\documentclass[%
 reprint,
superscriptaddress,
 amsmath,amssymb,
 aps,
prb,
]{revtex4-2}

\usepackage{amsmath}
\usepackage{graphicx,epstopdf}
\usepackage{dcolumn}
\usepackage{bm}
\usepackage{xcolor}
\usepackage{float}
\usepackage{soul}
\usepackage{enumerate}
\usepackage{natbib}
\usepackage{multirow}
\usepackage{bbold}

\begin{document}
\title{Slip-dominated  structural   transitions}

\author{Kanka Ghosh}
\email{kanka.ghosh@lspm.cnrs.fr}
\affiliation{CNRS, LSPM UPR 3407, Université Sorbonne Paris Nord, 93430 Villetaneuse, France}

\author{O\u{g}uz Umut Salman}%
 \affiliation{CNRS, LSPM UPR 3407, Université Sorbonne Paris Nord, 93430 Villetaneuse, France}
\author{Sylvain Queyreau}%
\affiliation{CNRS, LSPM UPR 3407, Université Sorbonne Paris Nord, 93430 Villetaneuse, France}

\author{Lev Truskinovsky}%
\email{lev.truskinovsky@espci.fr}
 \affiliation{PMMH, CNRS UMR 7636 ESPCI ParisTech, 10 Rue Vauquelin,75005 Paris, France}


\begin{abstract}


  We use molecular dynamics to show that  plastic slip is  a crucial component of the transformation  mechanism  of   a square-to-triangular  structural transition.  The latter is   a stylized analog of many other  reconstructive phase transitions. To justify our conclusions we use an atomistically-informed mesoscopic representation of the field of lattice distortions in molecular dynamics simulations.  Our approach reveals  a hidden alternating slip distribution behind  the seemingly homogeneous  product phase which points to the fact  that lattice invariant shears play a central  role  in  this class of phase  transformations. While the underlying pattern of anti-parallel displacements may be also  interpreted as microscopic shuffling, its precise crystallographic nature strongly suggests the plasticity-centered interpretation.

\end{abstract}

\maketitle

\section{Introduction}

Reconstructive phase transitions are the most widespread type of structural transformations in solids. These  transitions lack  the simplifying  group-subgroup relationship and  therefore cannot be described by the conventional Landau theory. The development of the equally encompassing  theory of reconstructive transitions is still   a challenge given that they involve breaking of chemical bonds and are characterized by micro-inhomogeneous configurations with slip, twinning, and stacking faults apparently intertwined \cite{toledano1996recon,  bhattacharya2004crystal, cahn1977symmetry, gao2016group, gao2017universal, otsuka2005physical, li2016assembly}.


The  BCC-HCP reconstructive  transition is  one of the most representative \cite{bancroft1956polymorphism, caspersen2004importance, kalantar2005direct, baruffi2022atomistic,young1991phase, banerjee2010phase, zong2020nucleation, grimvall2012lattice}. Its mechanism was  proposed by Burgers based on crystallographic analysis \cite{burgers1934process, mao1967effect, srinivasan2002mechanism, bassett1987mechanism, wang2013microstructural, masuda2004hcp, merkel2020microstructural, lee2013intersecting}. A salient feature of the Burgers mechanism  is an interplay between  a homogeneous shear and a superimposed alternating shuffling. The latter was  interpreted as  anti-parallel shifting of atomic layers which emerged as a result of softening of a finite wavelength phonon. However, the origin of such softening could not  be addressed   based on crystallography only and various attempts to  interpret it included references to structural mechanics,  energetics and kinetics \cite{mao1967effect, banerjee1998substructure, zhao2000thermal, cayron2016angular, cayron2010mechanisms, van2006experimental, zong2020nucleation, zahiri2023twinning,sanati2001landau,liu2009bcc,riffet2020role}. Similar problem exists for  the FCC-HCP reconstructive phase transition  which  can be accomplished crystallographically by the   coordinated  anti-parallel gliding of  Shockley partials  on every second close-packed crystallographic plane \cite{jin2005crystallography, olson1976general}. For both of these iconic reconstructive transformations the fundamental raison d'être for  the corresponding   antagonistic  displacements still remains obscure  \cite{gartstein1979fcc,fujita1972stacking,waitz1997fcc,liu2004stress,yang2006factors,liu2005thermally,li2017mechanism,zhang2022atomic,guo2023atomic,
li2025atomic,zhang2024cooperative,farabi2024fcc,freville2023comparison,rosa2022martensitic,
jin2005crystallography,galindo2024mathematical}, 
which is disappointing given that  the emerging pattern of anti-parallel, crystallographically specific, nanoscale  coordinated displacements is exactly  the  distinguishing feature of reconstructive transitions   which  places them outside   the classical Landau picture \cite{dmitriev1988definition,toledano1996recon,dmitriev2023discontinuous}.


In this paper we propose a novel general  interpretation of the shuffling phenomenon. Our conclusions are 
backed by the  systematic molecular dynamics (MD)  studies of  a prototypical model which unambiguously links the apparent shuffling  with highly cooperative plastic slip.  Specifically, we study the simplest   transition between  2D square (S) and  triangular (T) lattices \cite{arbib2023ericksen, conti2004variational, bhattacharya2004crystal}.  While this   square-to-hexagonal reconstructive phase change is of interest by itself \cite{e1980solidlike, eskildsen1997observation, keimer1994vortex, chang1998interpretation, holz1980defect, glasser1992energies, bayot1996giant, rao1997shape, hatch2001systematics, shirahata2012structural, cheng2013transition, qi2015nonclassical, peng2015two, peng2017diffusive, peng2023situ}, it can be considered  as a   stylized,  low dimensional, Bravais lattice based, analog of both emblematic BCC-HCP and FCC-HCP reconstructive transitions \cite{gao2020cayley, gao2019deformation, denoual2016phase}.

To reveal the underlying plastic slip in  our MD simulations, we shift attention from the conventional description of the transformation history in terms of \emph{individual}  atoms  to the  novel description  in terms of  evolving  atomic  \emph{neighborhoods}. This way of post-processing of MD data  allows us to   map  the transformation path  into the configurational space of the mesoscopic metric tensors. The  purely geometrical periodic tessellation  of the latter creates the  possibility to distinguish unambiguously between elastic and plastic deformations \cite{conti2004variational, bhattacharya2004crystal,baggio2019landau}. The application of  such atomistically-informed   representation of  lattice distortions  in the case of S-T transition reveals  that its  fundamentally non-affine mechanism involves  alternating lattice invariant   shears. We show that the crystallographically specific nature of such shears  points towards  a   plasticity-centered interpretation of the underlying  reconstructive transition.

 To corroborate the results of the  MD numerical experiments, we  also performed  parallel studies  of an athermal  molecular statics model  and a coarse grained mesoscopic model which directly deals with the evolution of atomic neighborhoods \cite{baggio2019landau, baggio2023homogeneous, salman2021discontinuous, baggio2023inelastic, perchikov2024quantized}.  The obtained qualitative agreement among all three models,  which differ in their  microscopic details, points to the possibility that the  proposed  slip-related  interpretation of the mechanism of the S-T transition is a robust feature of a broad   class of reconstructive transformations.
 
 The rest of the paper is organized as follows. In Section 2 we formulate the proposed approach to  post-processing of  the instantaneous MD  data in terms of the original nearest neighbors and show its advantages vis-à-vis the conventional interpretations in terms of the current nearest neighbors. In Section 3 we formulate our MD model and   present the result of numerical experiments by projecting them into the space of metric tensors. The supporting numerical experiments performed in the framework of athermal molecular statics and mesoscopic coarse grained models are discussed in Sections 4 and Section 5, respectively. In Section 6 we discuss our results vis-à-vis some previous modeling work in 2D and mention some parallels with what is known about BCC-HCP and FCC-HCP reconstructive transitions in 3D.  Finally, our conclusions are summarized in  Section 7.



\section{Preliminaries}

We first discuss  the proposed novel approach   to the  interpretation of   the results of MD simulations. The main idea  is to use  individual atomic position data   to extract  the local values of the deformation gradients.
This amounts  to post-processing  the instantaneous MD  data which are then  interpreted as representing  piecewise linear strain fields \cite{inoue1995molecular, Kruyt1996micromechanical, sengupta2000elastic,mott1992atomic, ward2006mechanical}. 

\begin{figure}[H]
\centering
\includegraphics[width=0.4 \textwidth]{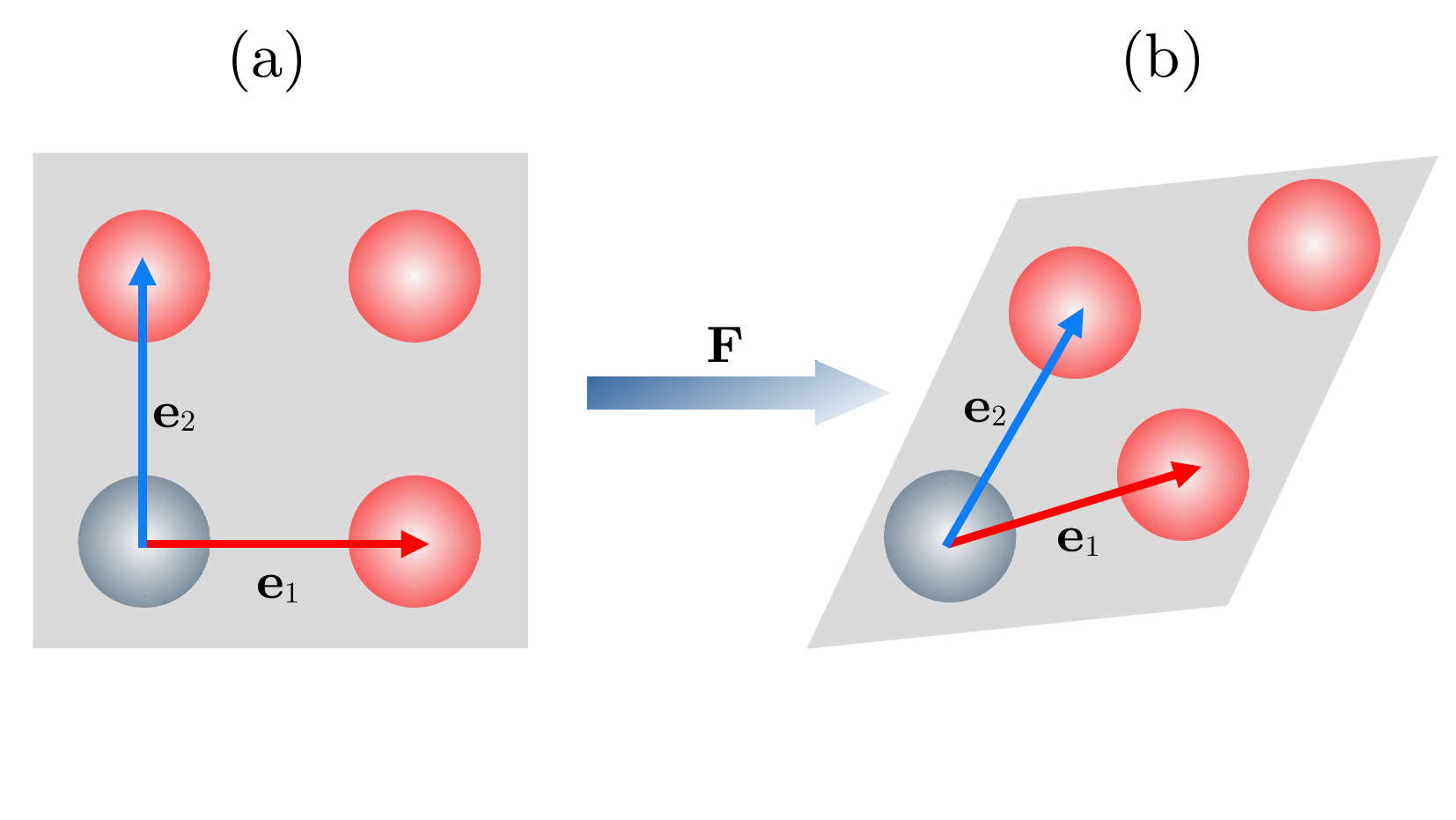}
\caption{The schematic description of the deformation of the chosen  `atomic neighborhood' ; (a), (b) show the reference and the deformed states, respectively.}
\label{fig:1}
\end{figure}

Indeed, by denoting the reference discrete atomic positions   $\textbf{x}$ and the deformed atomic positions   $\textbf{y}$, we can define the \emph{approximate}  deformation gradient  $\textbf{F}$ by minimizing the   error function \begin{equation} \label{F1}
\sum  \parallel \Delta \textbf{y} - \textbf{F} \Delta \textbf{x} \parallel^2
\end{equation}
 with  summation  over the pairs of elements inside the chosen neighborhood of a  given site  \cite{falk1998dynamics, falk1999molecular, horstemeyer1999strain, mitchell2016strain, sartori2023evolutionary, eckmann2019colloquium,zimmerman2009deformation}. 
Suppose first that in a  two-dimensional lattice  a reference point is represented by a vector $\textbf{x}$ = \{x$_{1}$, x$_{2}$\} while  its known  deformed position is represented by the vector $\textbf{y}$ = \{y$_{1}$, y$_{2}$\}. Then, the \emph{exact} value of the deformation gradient  
   \begin{equation} \label{F}
    F_{iI} =  \frac{\partial y_i}{\partial x_I},
   \end{equation}
known as well; note that in \eqref{F}  the indexes  $i$, $I$ refer to deformed and reference coordinate systems, respectively. Now, if in MD simulations $\textbf{R}^{\alpha\beta}$ and $\textbf{r}^{\alpha\beta}$ are the vectors connecting  atom  $\alpha$ with its neighbor $\beta$ in the reference and in the actual configurations,   the  approximate   deformation gradient can be obtained by minimizing mean-square difference \eqref{F1} between the actual displacements of the neighboring atoms relative to the chosen central atom and the relative displacements  that they would have had if they were in a region of uniform strain. In Fig. \ref{fig:1}   we   schematically  show a   deformation of an  atomic neighborhood. Given that we deal with weakly distorted lattices   the   sampling neighborhood is  limited here to two complementary triangular domains.  In other words, as a representative  atomic neighborhood we have chosen  two non collinear nearest neighbors and one of the second nearest neighbors.  In general, the result of the proposed approximation procedure can be written   in the form \cite{zimmerman2009deformation}
   \begin{equation}
    F_{iI}^{\alpha} = \omega_{iM}^{\alpha} \left(\eta_{IM}^{\alpha}\right)^{-1},
   \end{equation}

where 
   \begin{equation}
    \omega_{iM}^{\alpha} \equiv \sum_{\beta = 1}^{n} r_{i}^{\alpha\beta}R_{M}^{\alpha\beta},\,\,
\eta_{IM}^{\alpha} \equiv \sum_{\beta = 1}^{n} R_{I}^{\alpha\beta}R_{M}^{\alpha\beta}.
   \end{equation}   
After the approximate deformation gradient   
 \begin{equation}
  \textbf{F} =  \partial \textbf{y}/\partial \textbf{x}
     \end{equation}
is recovered, we can compute 
the atomistic metric  tensor  
       \begin{equation}
        \textbf{C} = \textbf{F}^{T}\textbf{F}.  
          \end{equation}  The possibility of mapping the results of MD simulations into  the $\mathbf{C}$-space   is of great interest because its crucial  subspace, 
\begin{equation}
\det(\textbf{C})= 1, 
\end{equation}
 is  naturally tessellated by the action of the global symmetry group of   Bravais lattices. The latter can be viewed as a  finite strain extension of the crystallographic point group.   In the case of interest,  such group has  the well known matrix representation  
    \begin{equation}
\textit{GL}(2, \mathbb{Z}) = \{\mathbf{m}, \ m_{IJ} \in \mathbb{Z}, \ \det(\mathbf{m}) = \pm 1\},
   \end{equation}
    see for instance    \cite{ericksen1970nonlinear, ericksen1977special, ericksen1980some, boyer1989magic, folkins1991functions, wang1993crystal, waal1990general, parry1976elasticity, parry1977crystallographic, parry1998low, pitteri1984reconciliation, pitteri2002continuum, kaxiras1994energetics, engel2012geometric, michel2001fundamental,bhattacharya2004crystal, conti2004variational}. 
More specifically, we can   visualize  in this case the  subspace of metric tensors  with coordinates $(C_{11}, C_{22}, C_{12})$ describing  isochoric deformations
 \begin{equation}
  C_{11}C_{22} -C_{12}^2=1
   \end{equation}  
using a stereographic projection of the corresponding hyperbolic surface onto a unit (Poincaré ) disk.  To obtain such a representation in the case when the reference state is a square lattice, we first define the corresponding reference basis   ${\bf e}_1=\{1,0\}, {\bf e}_2=\{0,1\}$. The deformed basis is   ${\bf f}_i={\bf F}{\bf e}_i$, where $i=1,2$ and ${\bf F}$ is the deformation gradient. Under the assumption that  $\det {\bf F} = 1$ the  metric tensors  $\bf C=\bf F^T\bf F$  can  be then projected onto the Poincaré disk  using the rectangular coordinates  
      \begin{equation}
 x=t(C_{11}-C_{22})/2,\,\, y=t C_{12},
    \end{equation}

\begin{figure}[H]
\includegraphics[width=0.5 \textwidth]{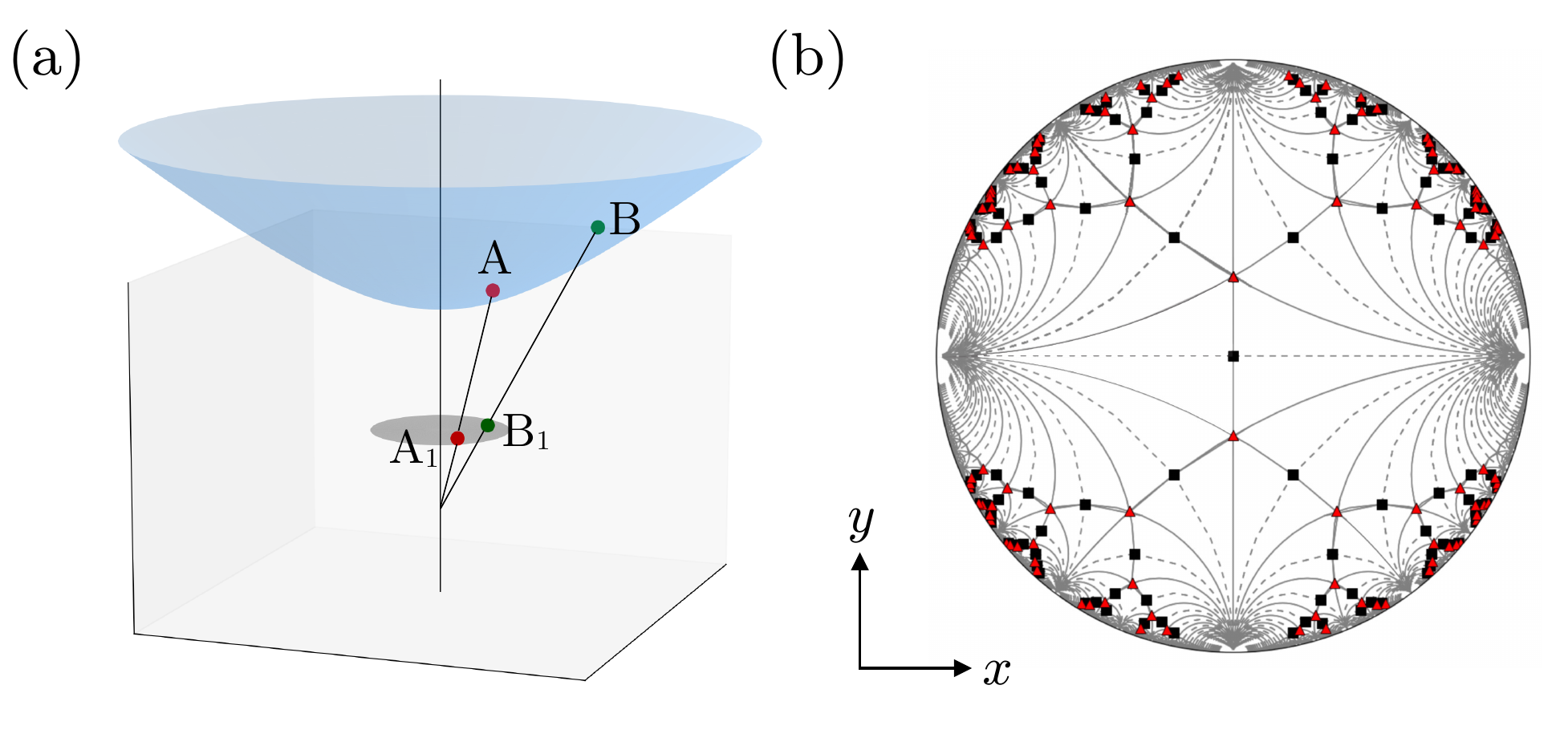}
\caption{(a) 3D hyperbolic surface $C_{11}C_{22} -C_{12}^2=1$ in the configurational space of metric tensors $C_{11}, C_{22}, C_{12}$ projected onto a Poincaré disk: the points A, B  on such  a surface are mapped to the points A$_1$, B$_1$ on the disk. (b) Poincaré disk: thin lines   indicate the boundaries of minimal periodicity domains;  points on the disk describing equivalent square and triangular lattices  are marked  by black squares and red triangles, respectively.}
\label{fig:2}
\end{figure}

where $t= 2(2+C_{11}+C_{22})^{-1}$. This mapping is  schematically illustrated in  Fig. \ref{fig:2}(a).

\begin{figure}[H]
\centering
\includegraphics[width=0.48\textwidth]{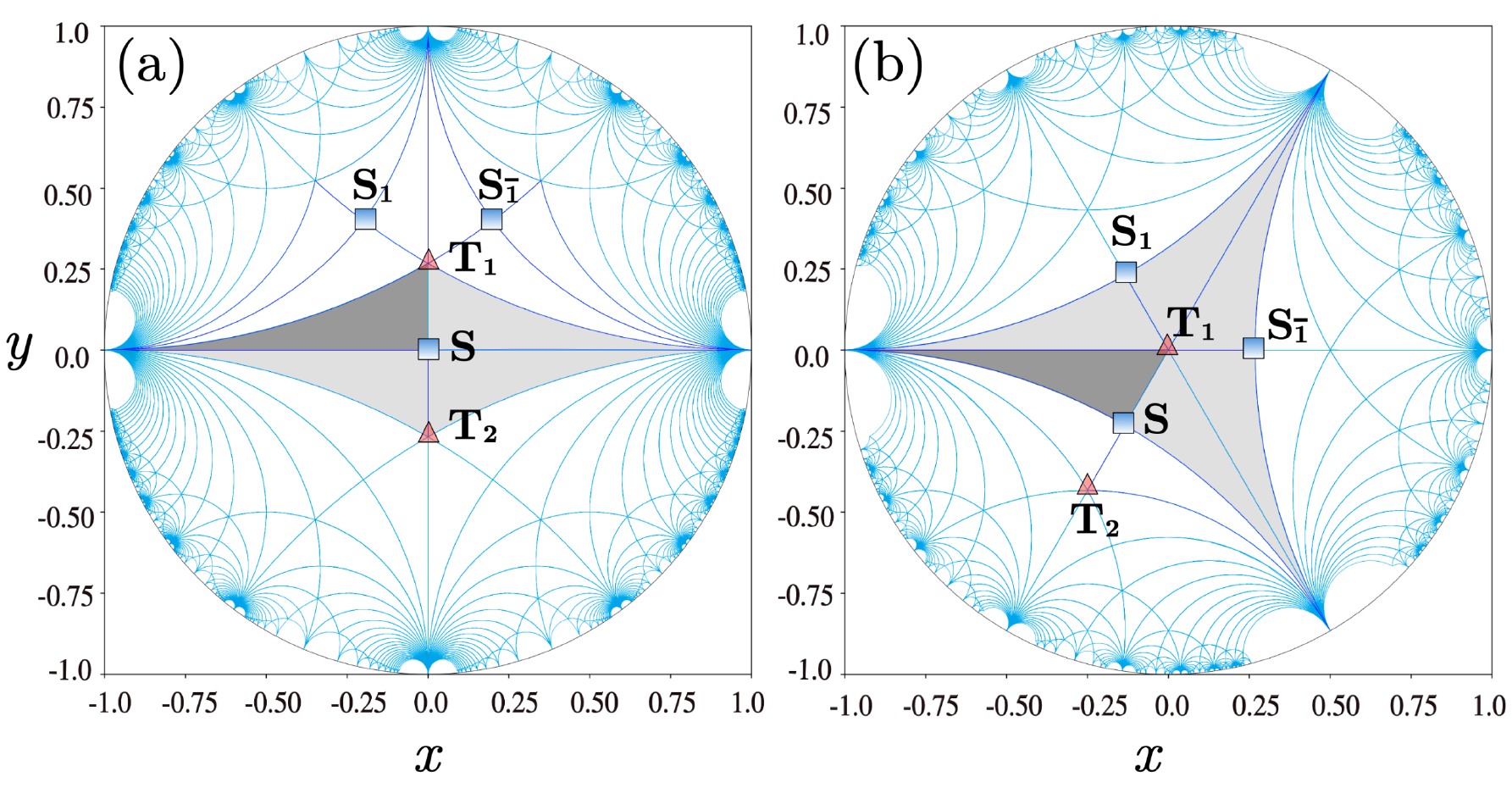}
\caption{Stereographic projection on a Poincare disk of the configurational space of metric tensors $\mathbf{C}$  with $\det{\mathbf{C}}=1$. In (a,b) the  reference states are the square phase  S   and the triangular phase  T$_1$, respectively. Dark gray indicates the minimal periodicity domain, light gray --  the minimal elastic domain; blue lines represent the tessellation induced by the $\textit{GL}(2, \mathbb{Z})$ global symmetry. Two additional square variants S$_1$, S$_{\bar{1}}$,  accessible from T$_1$ are also shown.} 
\label{fig:3}
\end{figure}

As we have already mentioned, he action  of the infinite discrete symmetry  group $\textit{GL}(2, \mathbb{Z})$
divides the Poincaré disk into periodicity domains, see Fig.   \ref{fig:2}(b).
 The minimal periodicity  domain  of this kind,  known as the   fundamental domain, can be represented in our case explicitly
\begin{equation}
\mathcal{D}= \{ 2C_{12}  \le  \min(C_{11},C_{22}) \},
\end{equation}
 see the dark gray area in Fig. \ref{fig:3}(a).  It corresponds to the `minimal' choice for the lattice vectors $\tilde{\mathbf{e}}_1, \tilde{\mathbf{e}}_2$, which can be specified using the classical Lagrange reduction algorithm \cite{conti2004variational,  parry1976elasticity, parry1977crystallographic, parry1998low, pitteri1984reconciliation, pitteri2002continuum,  engel2012geometric, michel2001fundamental}. 

The three boundaries of the  fundamental  domain   $\mathcal{D}$   

can be presented explicitly in the parametric form:
 \begin{equation} \label{1}
\mathbf{C} = \begin{pmatrix}
\alpha^2 & 0 \\
0 & \frac{1}{\alpha^2}
\end{pmatrix}, 0<\alpha\leq1,
\end{equation}
 \begin{equation} \label{2}
\mathbf{C} = \begin{pmatrix}
\alpha^2/4+1/\alpha^2 & -\alpha^2/4+1/\alpha^2 \\
-\alpha^2/4+1/\alpha^2 & \alpha^2/4+1/\alpha^2
\end{pmatrix}, 0<\alpha\leq\sqrt{2}
\end{equation}
and
 \begin{equation} \label{3}
\mathbf{C} = \begin{pmatrix}
\alpha^2 & \alpha^2/2 \\
 \alpha^2/2 &  \alpha^2/4+1/\alpha^2
\end{pmatrix},0<\alpha\leq \gamma.
\end{equation}
The   $\textit{GL}(2, \mathbb{Z})$ copies (replicas) of these   boundaries, constitute  the tessellation of the configurational space, represented in Fig. \ref{fig:2}(b) and Fig. \ref{fig:3}(a) by the thin black lines which are divided (artificially) in  Fig. \ref{fig:2}(b)  into  solid and dashed ones for easier identification, see   \cite{baggio2023inelastic, perchikov2024quantized}  for more details. 

  The   \emph{elastic   domain},   represents   the minimal set where the mechanical  response is elastic.   It is also known in the literature   as the maximal Ericksen-Pitteri neighborhood  \cite{pitteri1984reconciliation, pitteri2002continuum, conti2004variational}.  The    elastic   domain  can be generated  from  $\mathcal{D}$ by applying discrete mappings representing  the classical crystallographic point group $P(\mathbf{e_I})$  containing only rigid rotations and  
used to characterize  material symmetries within classical continuum elasticity \cite{pitteri2002continuum,conti2004variational}. The    elastic   domain  is identified  in  Fig. \ref{fig:3}(a)  by  light gray color.

%

Our  Fig. \ref{fig:3}(a,b)   provide equivalent information with the only difference that  in  Fig. \ref{fig:3}(a)   the  $\mathbf{C}$-space is centered  around  the reference square lattice (point S),   while  in Fig. \ref{fig:3}(b)    the reference  lattice is triangular (point T$_1$). In the latter case,  the construction of the  $\textit{GL}(2, \mathbb{Z})$ induced tessellation of the Poincaré disk and recovery of the corresponding   symmetry-induced periodicity structure in the space of metric tensors $\mathbf{C}$, proceeds through the following steps. We first represent the basis vectors of the triangular lattice  T$_1$ in the coordinates of the basis of the square lattice S to obtain: ${\bf h}_1=\{\gamma,0\}$ and ${\bf h}_2=\{\gamma/2,\gamma\sqrt{3}/2\}$, where $\gamma=(4/3)^{1/4}$.  We then introduce a matrix $ {\bf H} $  whose columns are the vectors ${\bf h}_{1,2}$. Since   ${\bf h}_j={\bf H} {\bf e}_i$ the metric tensor in this new (triangular lattice)  basis takes the form $ {\bf C'}  ={\bf H}^{-T}{\bf C}{\bf H}^{-1}$ where $ \bf C $ is the metric tensor in the square lattice basis. Next, the  components of the tensor $ {\bf C'}$ are  stereographically projected into the Poincaré disk using the same mapping as we used above  which gives rise to a tessellation presented in Fig. \ref{fig:3}(b). While  in both cases shown in  Fig. \ref{fig:3}(a,b)   the   fundamental  domain  $\mathcal{D}$ has the same triangular shape, the  elastic domains are different  reflecting the difference in the corresponding point groups. More specifically, the  point group involves four rotations when the reference lattice is square and six rotations when it is triangular, which is illustrated by  the different number  light gray areas in Fig. \ref{fig:3}(a,b). 

Even a simple juxtaposition  of the structure of the elastic domains  in  Fig. \ref{fig:3}(a) and  Fig. \ref{fig:3}(b) already suggests  some striking differences between the direct transition S-T and its inverse  T-S transition, if both are taking place in isotropic (unbiased) conditions with control parameters being either  temperature or pressure. Note first that the single square  variant  S  should necessarily transform to the mixture of two  triangular variants T$_1$  and    T$_2$. The latter are located in  Fig. \ref{fig:3}(a) exactly at the boundary of the elastic domain. However, our Fig. \ref{fig:3}(b) shows that,  in fact,  the corresponding  two triangular lattices (T$_1$  and    T$_2$)   belong to different  elastic domains which also follows from an  observation that  they are separated by a lattice invariant shear. The fact that such shear corresponds to an elementary plastic slip  provides  the first indication that the  mixture of the variants  T$_1$  and    T$_2$ emerging as a result of S-T transformation, should be considered as a plastically deformed triangular phase. Similarly, if we consider  an isotropic T-S  transformation originating at the state T$_1$, one can expect the product phase to be a  mixture of the three variants of the square lattice: S,  S$_1$, and  S$_{\bar{1}}$. Since, in the perspective of Fig. \ref{fig:3}(a),   all of them belong to different elastic neighborhoods and differ by lattice invariant slips, the corresponding product  square phase  will  be even more severely plastically deformed.


Additional insights can be obtained if we  also project    the  energy landscape
into  the $\mathbf{C}$-space. To construct such a landscape it is sufficient  to  apply  homogeneous deformation $\mathbf{C}$   to a sufficiently large  set of atoms, while accounting for all  microscopic  interactions.  One can then  use the Cauchy-Born rule \cite{ericksen2008on,ming2007cauchy} and write 
\begin{equation} \label{V1}
    \phi(\mathbf{C}) = \frac{1}{2\Omega} \sum_{\mathbf{x}}\sum_{\mathbf{x}^{c}\in \mathcal{N}(\mathbf{x})} V(\sqrt{R_{i}C_{ij}R_{j}}),
 \end{equation}
where $V(r)$ is a  pairwise  interaction potential,  $R_i$ are the vectors representing reference points  and the internal summations extend over  all points $\mathbf{x}^{c}$ belonging to the cut-off neighborhood $\mathcal{N}(\mathbf{x})$. 
In our work  to ensure that  a square lattice is the ground state  we used the potential in the form  \cite{boyer1996}
\begin{equation} \label{V}
 V(r) =  a/r^{12}  - c_{1} \exp{[-b_{1}(r-r_{1})^{2}]} - c_{2} \exp{[-b_{2}(r-r_{2})^{2}]}.
 \end{equation}
Here  $r_1$ is the lattice constant and $r_2$ is the second nearest neighbor distance and we  used   the parameter values $c_1=c_2=2$  and  $b_1=b_2=8$;  the choice  $r_2$/$r_1$=1.425 produced  the  square  ground state  with   a lattice constant  equal to 1.0659 \AA{}. 

Note  that it is sufficient to use  \eqref{V1} inside   the minimal periodicity  domain $\mathcal{D}$.  The globally symmetric potential  $\phi(\mathbf{C})$  can be then extended to the whole $\mathbf{C}$-space using the  $\textit{GL}(2,\mathbb{Z})$   periodicity.  We recall that it implies the mapping of an arbitrary metric tensor $\mathbf{C}$ into the domain $D$  which produces its Lagrange-reduced image  $\tilde{\mathbf{C}}=\mathbf{m}^{T} \mathbf{C m}$ where $\mathbf{m}$   is a unimodular integer-valued   matrix. Since  $\tilde{\mathbf{C}}$  is an image of  $\mathbf{C}$ inside the minimal periodicity domain,  we can compute the value of the energy $\phi(\mathbf{C})$ by simply applying the equality  $\phi(\mathbf{C})=\phi(\tilde{\mathbf{C}})$ where $\phi(\tilde{\mathbf{C}})$ is obtained  by  our Cauchy-Born recipe \eqref{V1}. 
\begin{figure}[H]
\centering
\includegraphics[width=0.4\textwidth]{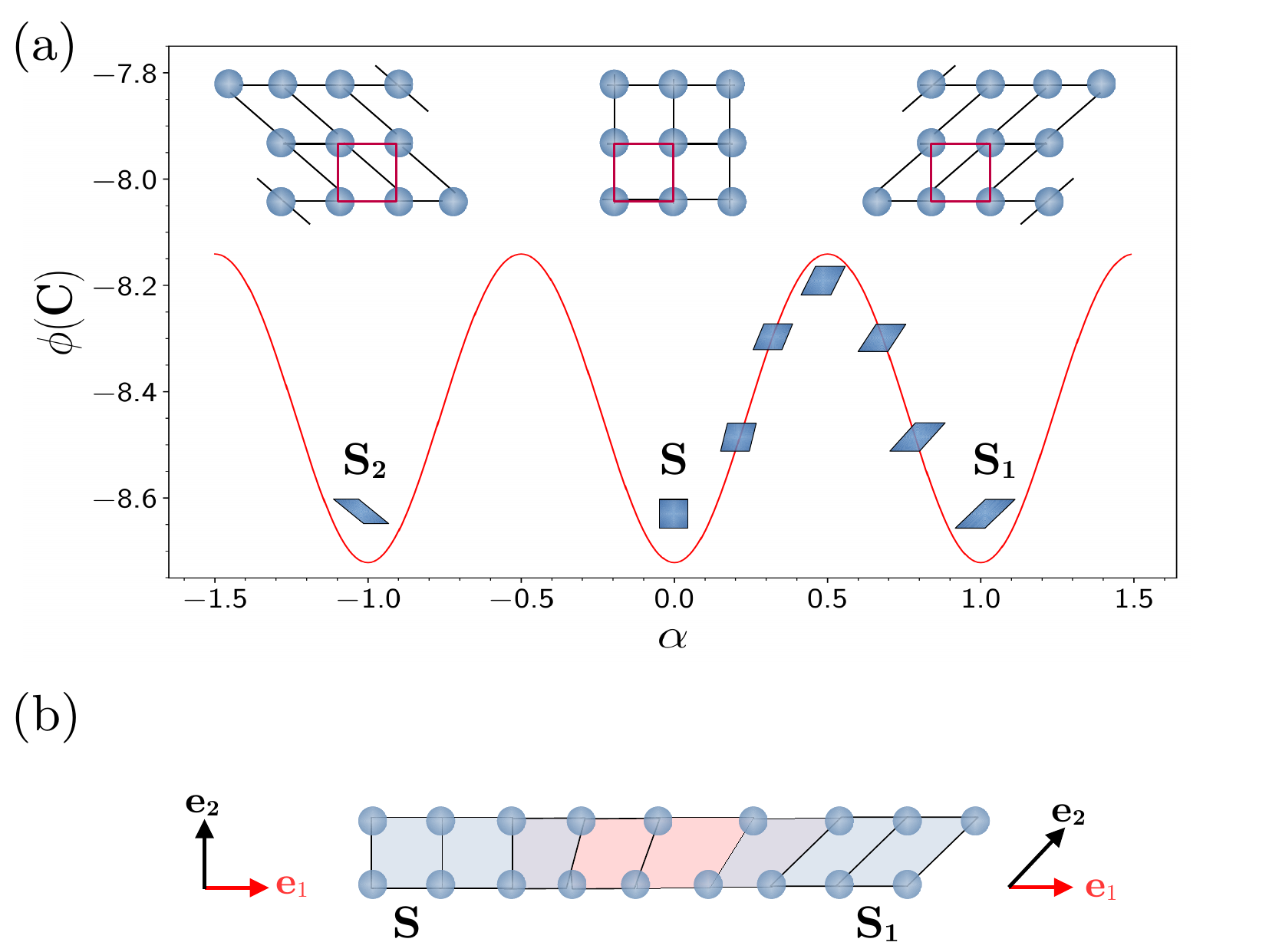}
\caption{(a) Schematic representation of the strain energy   $\phi (\mathbf{C})$ as a function of a simple shear amplitude $\alpha$ (see the text). The insets schematically depict the equivalent square lattice configurations S$_2$, S and S$_1$. (b) Schematic presentation of an edge dislocation.}
\label{fig:4}
\end{figure}

Before we discuss the resulting energy landscape $\phi(\mathbf{C})$, it is instructive  to  schematically  illustrate the structure of such a landscape around  lattice invariant shears, which are  crystallographically specific affine volume preserving deformations that  map  an infinite lattice into itself. We recall that the general lattice invariant shears are described by  the  deformation gradients represented by the integer valued unimodular matrices  
$\mathbf{m}$, see \cite{baggio2019landau, salman2021discontinuous} for the  details.  To this end  we  can start with  a square lattice (S)  and deform it   homogeneously applying    the simple shear
\begin{equation} \label{alpha}
\mathbf{F}(\alpha) = \mathbb{1} + \alpha \mathbf{e}_1 \otimes \mathbf{e}_2,
\end{equation}
 where   $\alpha$ denotes the amplitude of shear, while $\mathbf{e}_1$, $\mathbf{e}_2$ represent unit base vectors along $x$ and $y$ directions, respectively. 

Our Fig. \ref{fig:4}(a) presents a schematic  structure  of the strain energy landscape  $\phi (\mathbf{C})$  along such a one parametric family of homogeneous deformations.   
 As  we increase the parameter  $\alpha$  from zero, which corresponds to a minimum of  the energy, the function $\phi (\alpha)$  first  reaches its  maximum but then decreases  reaching again  a symmetry related minimum at $\alpha$ = 1. At this point  an equivalent lattice is obtained. Similarly, negative increment of $\alpha$  transforms  at $\alpha$ = -1  the original lattice configuration  into yet another symmetric configuration with exactly the same energy. The insets of Fig. \ref{fig:4}(a) schematically describe three equivalent lattice configuration S, S$_1$ and S$_2$ corresponding to  three equivalent energy minima. Note that if we remove the bonds in those insets and leave only atoms, the  corresponding three atomic configurations will be indistinguishable. Furthermore, our Fig. \ref{fig:4}(b) presents a schematic structure of an edge dislocation which can be viewed as a `domain wall' between a sheared (S$_1$ in this case) and undeformed (S) `phases'.

\begin{figure}[H]
\centering
\includegraphics[width=0.45 \textwidth]{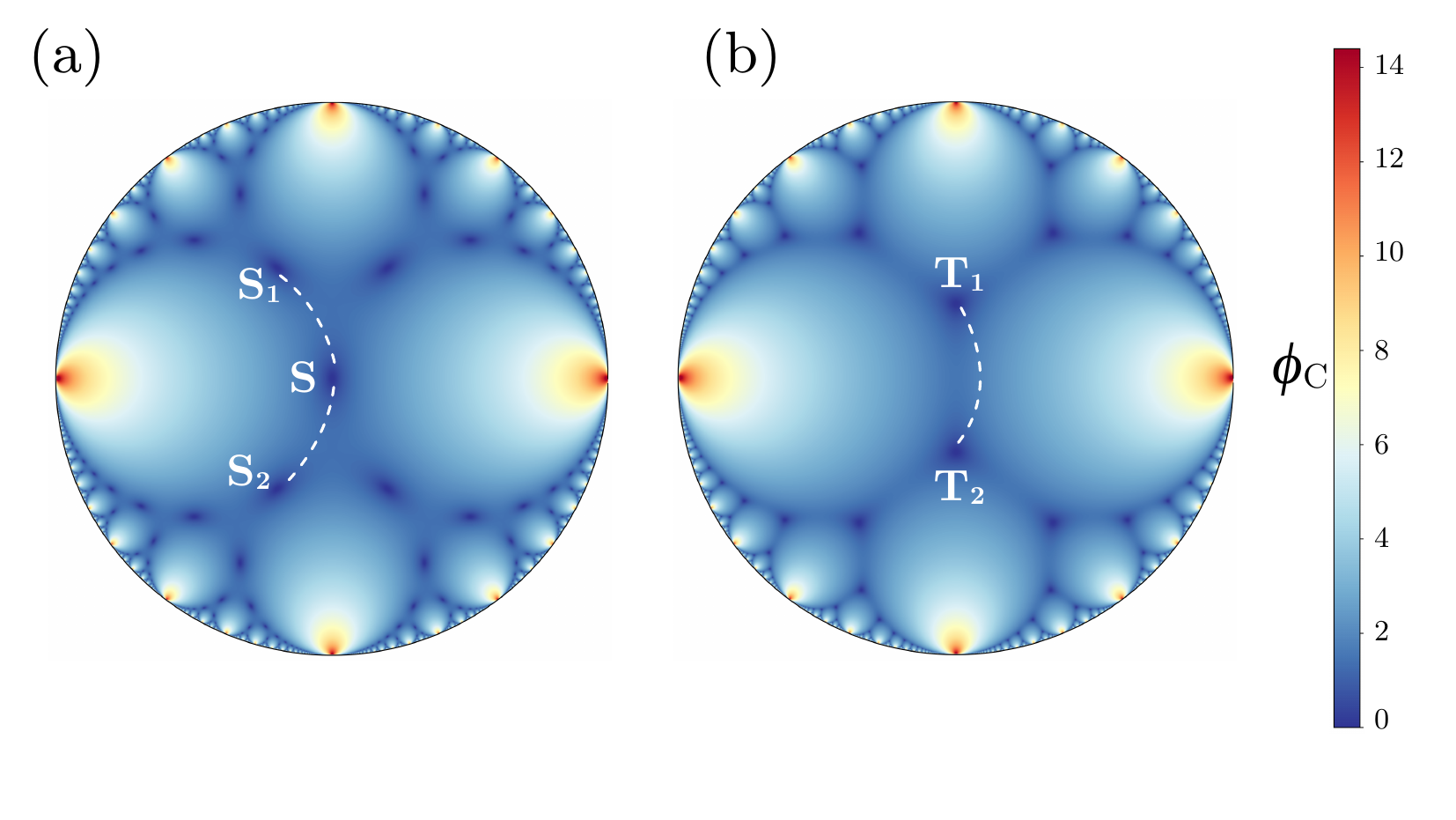}
\caption{(a) Schematic presentation of an edge  dislocation linking equivalent square lattices  on a Poincaré disk. (b)  Similar edge dislocation in the case of triangular lattice. Dashed lines represent examples of lattice invariant simple shear paths. The colors  represent  the  strain energy level.}
\label{fig:5}
\end{figure}
 
The complete  $\textit{GL}(2, \mathbb{Z})$ invariant energy landscape,  emerging if we apply the    Cauchy-Born rule,  is shown in Fig. \ref{fig:5} (a,b).  The parameters are chosen to ensure that the  square lattice represent  the ground state in Fig. \ref{fig:5} (a) while  the triangular lattice is the ground state in Fig. \ref{fig:5} (b).   In both cases the  simple shear paths connecting equivalent zero energy lattice configurations are shown by white dashed curves. Thus, the equivalent square lattices S, S$_1$ and S$_2$  are linked  in Fig. \ref{fig:5} (a)  via  the simple shear path  \eqref {alpha}.  Similarly, the equivalent triangular lattices T$_1$ and T$_2$ can be linked  via  two symmetric  simple shear paths   only one of which is  shown in Fig. \ref{fig:5}(b). These paths  can be viewed as  a rough representation  on the corresponding Poincaré disks of  the  cores of the associated   edge dislocations.

\section{Molecular dynamics}

We reiterate that the goal of our MD numerical experiments was to reproduce  pressure induced  prototypical reconstructive phase transition from a square crystal phase (S, plane space group  p4mm) to a close-packed hexagonal lattice crystal phase, interpreted  here as triangular lattice  (T, plane space group p6mm). The particle interaction potential was chosen in the form  \eqref{V} where  we used the fixed ratio $r_2$/$r_1$=1.425 to ensure that  at zero pressure a square lattice  with a lattice constant  equal to  1.0659 \AA{}  is in the ground state.  We  used a cutoff  distance $r_c=2.5 \AA{}$ where the potential was smoothly reduced to zero. All the molecular dynamics simulations were carried out using LAMMPS which includes  velocity Verlet algorithm  \cite{thompson2022lammps, Lammps}.

We simulated $10^4$ atoms with  periodic boundary conditions (PBC) and typically  performed 10$^7$ MD steps (= 1 ns) in each run.  First the square crystal was  equilibrated within NVT ensemble with 10$^5$ time  step sizes each equal to   $\Delta t$  = 0.0001 ps. The  pressure control protocol was  implemented  within isothermal-isobaric (NPT) ensemble till the square phase was  marginalized. As 

\begin{figure}[H]
\centering
\includegraphics[width=0.5\textwidth]{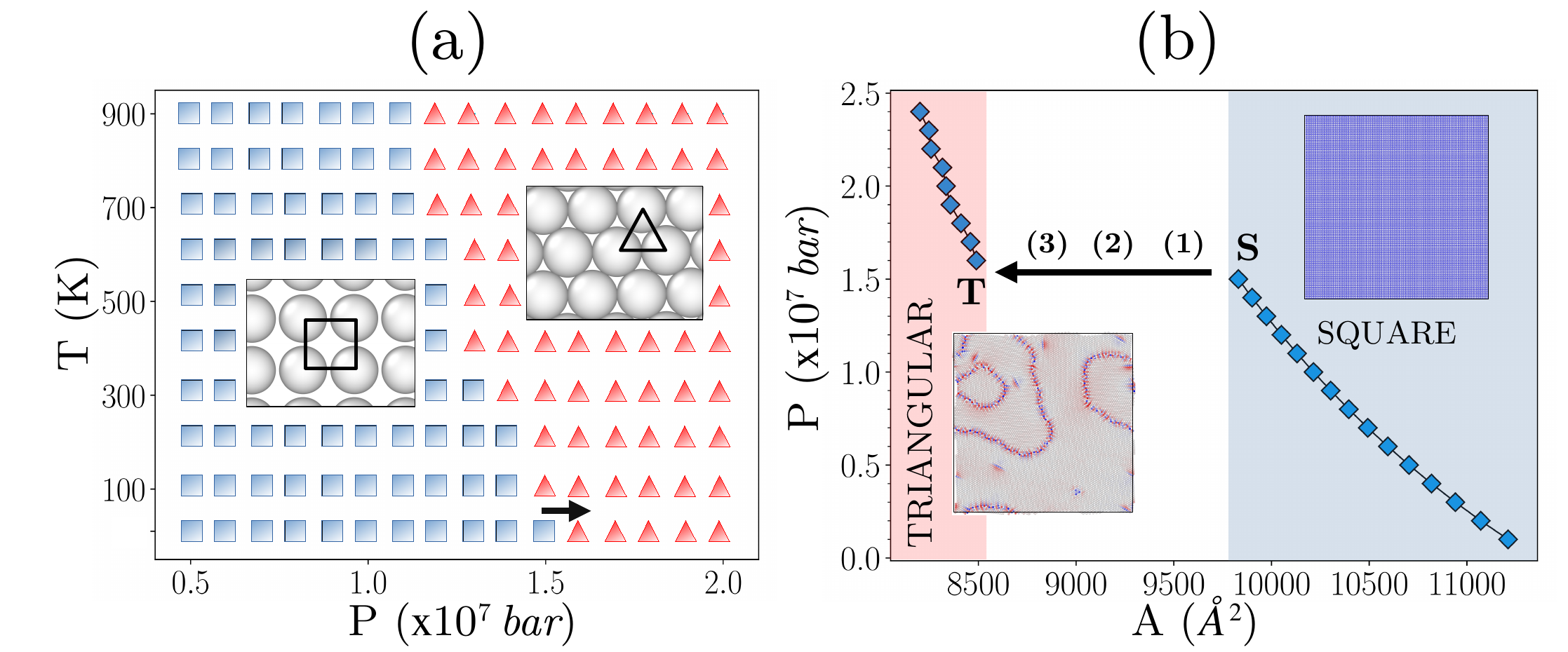}
\caption{  MD simulated   square-to-triangular transition: (a)  kinetic T-P  phase diagram  showing only the direct  transition; (b) the same  transition in P-V (area A) performed at T = 10 K (as indicated in (a) by an arrow).}
\label{fig:6}
\end{figure}

\noindent typically done  in NPT ensemble, it was the  volume (area A in our 2D system) of the simulation domain that was changed to reach the target pressure \cite{frenkel2002understanding}.  To construct phase diagram,  a broad  range of temperatures (10 K - 900 K) and pressures (0.1 $\times$ 10$^7$ bar - 2.4 $\times$ 10$^7$ bar) was covered.   Instead, to study  microstructure formation   we fixed temperature at  T = 10 K.



\begin{figure}[H]
\centering
\includegraphics[width=0.42 \textwidth]{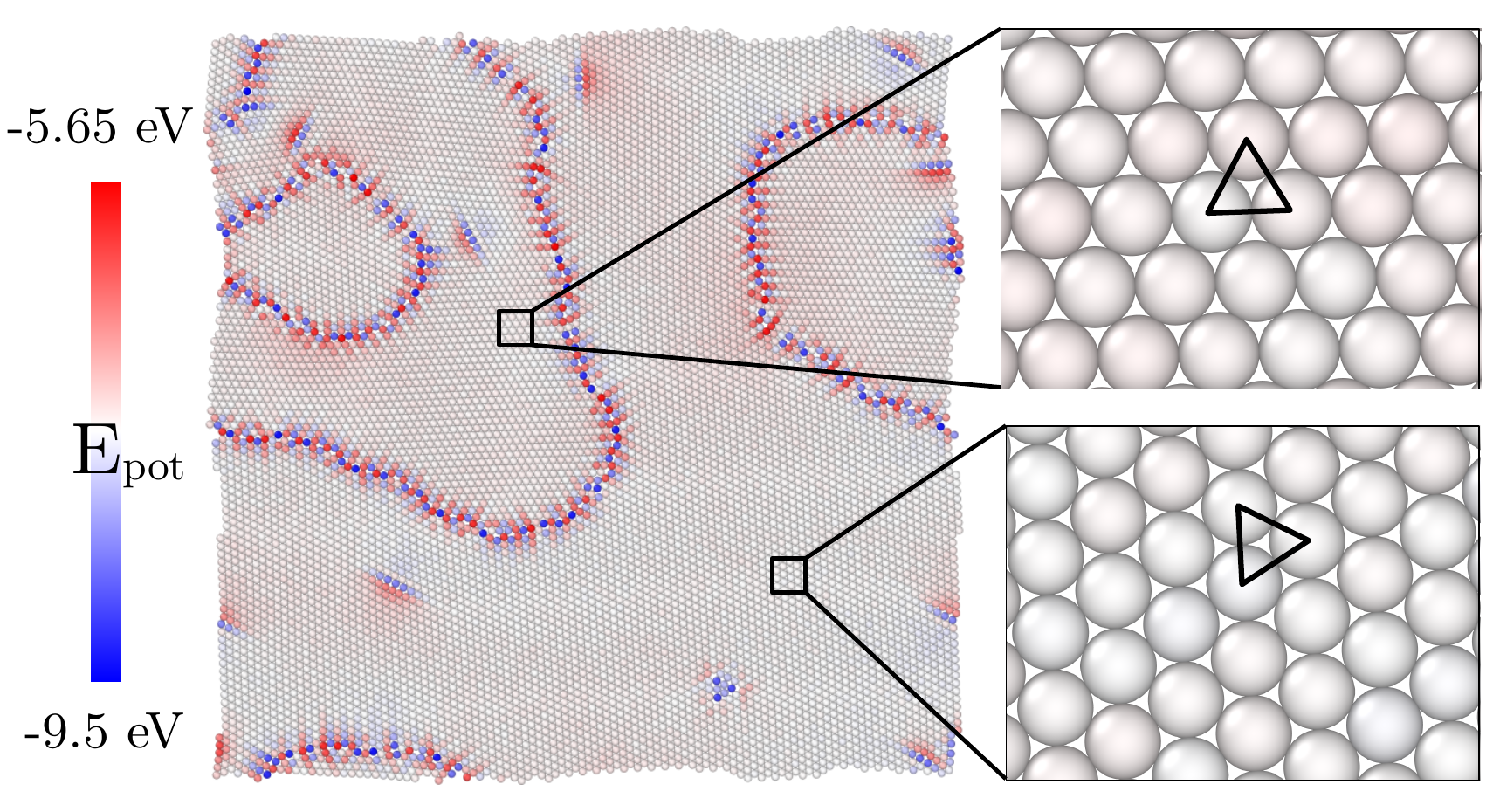}
\caption{Multigrain configuration of transformed triangular lattice (T) colored according to the potential energy of each atom. The atomic structure of two triangular grains with 30$^\circ$ misorientation is illustrated  in the insets.}
\label{fig:7}
\end{figure}

\begin{figure}[H]
\centering
\includegraphics[width=0.4\textwidth]{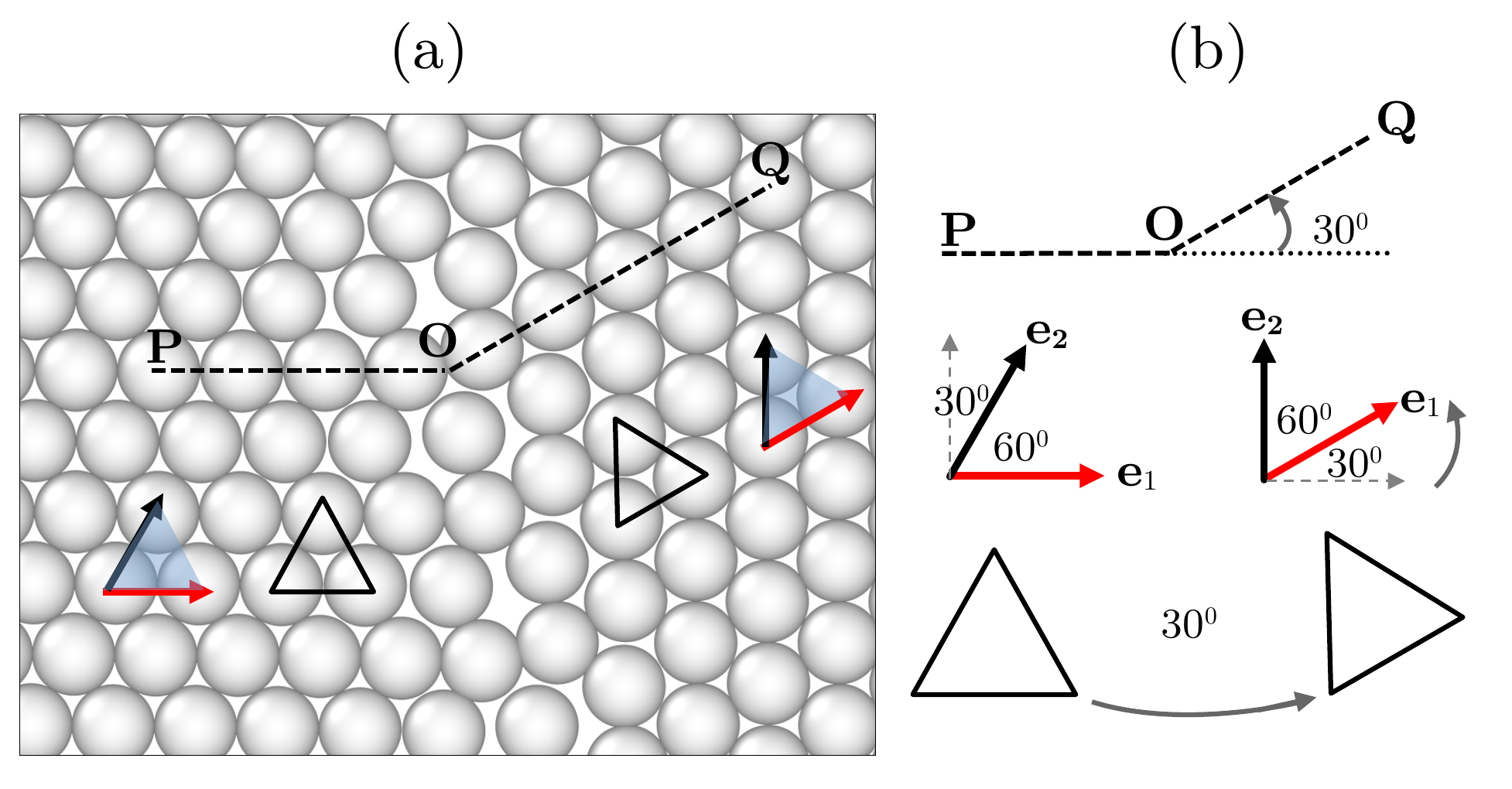}
\caption{(a) A fragment of the product polycrystalline triangular phase  showing two misoriented grains separated by a grain boundary; (b) basis vector  $\mathbf{e}_1$ and $\mathbf{e}_2$ in both  grains.      
}
\label{fig:8}
\end{figure}

The obtained   (kinetic) T-P  phase diagram  addressing only  the direct S-T transition is  shown in Fig. \ref{fig:6}(a). The predicted negative slope of the   stability/coexistence curve   agrees with similar numerical experiments \cite{damasceno2009pressure, lee1989mechanism, shuang2022atomistic} and is also consistent with the equilibrium data for BCC-HCP transformation in iron  \cite{dorogokupets2017thermodynamics, dewaele2015mechanism}.  In Fig. \ref{fig:6}(b), the same S-T  transition is shown in P-A coordinates at fixed  temperature $T = 10 K$. A salient feature of the observed phenomenon is that a  originally pure-crystalline  homogeneous square lattice   transforms into a highly inhomogeneous polycrystalline texture. This is illustrated in more detail in  Figure \ref{fig:7} where we zoom in into fragments of the product triangular lattice and show in the insets differently  oriented  hexagonal  grains.  The observed misorientation angles between the grains are  not arbitrary. Thus, our Fig. \ref{fig:8}(a) shows a fragment of the product triangular lattice featuring two grains  and one can see  that the basis vectors in one  grain need to be rotated

by 30$^\circ$ to transform into the basis vectors  of the other  grain. It is also  clear that the purely crystalline grains are separated  by  dislocation-rich  grain boundaries.

\begin{figure}[H]
\centering
\includegraphics[width=0.5\textwidth]{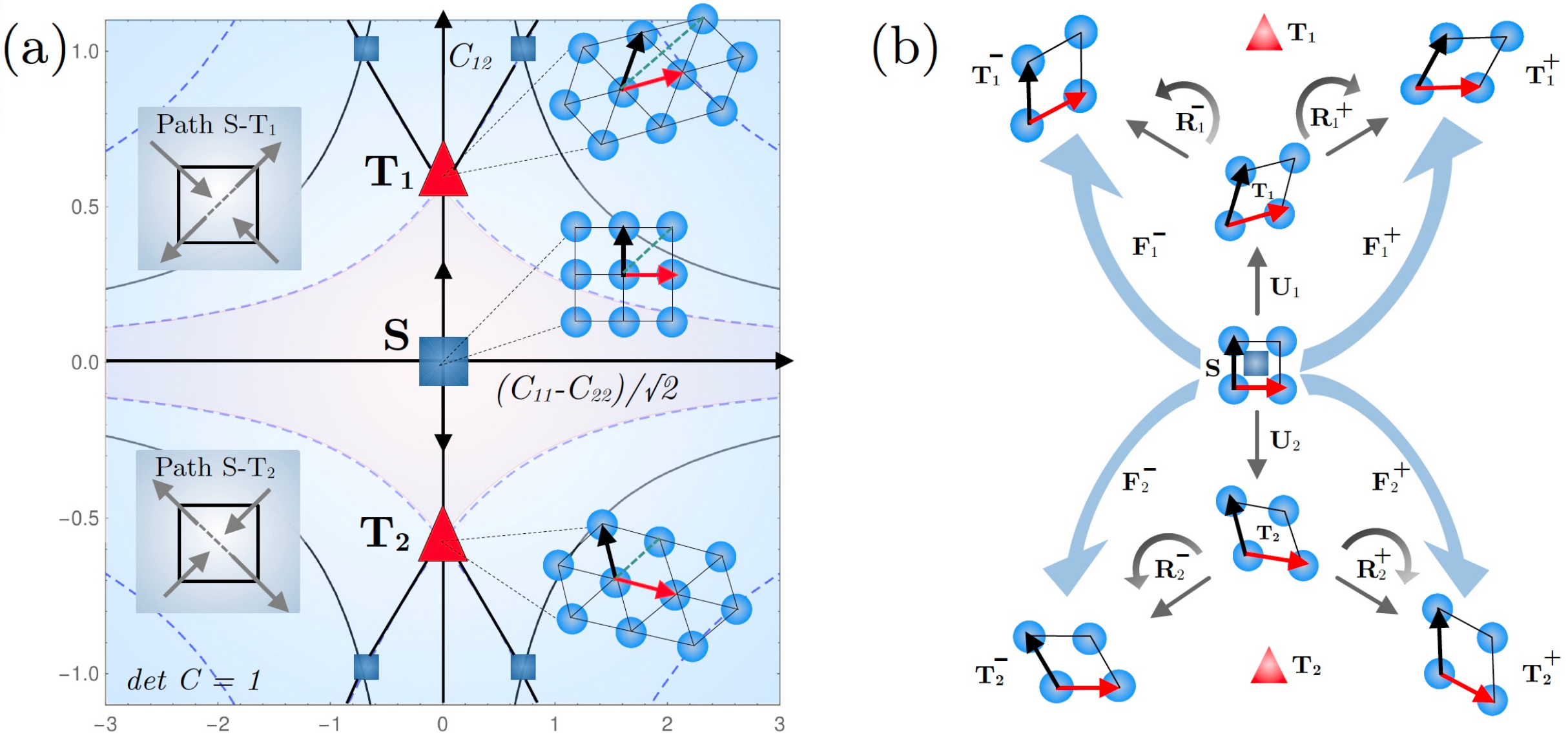}
\caption{(a) A fragment of the configurational space of metric tensors $\mathbf{C}$  showing the original square lattice (point S) and two equivalent versions of the triangular lattice (points T$_1$ and  T$_2$). Solid arrows are directed along the two symmetric pure shear (rhombic) paths S $\to$ T$_1$ and S $\to$ T$_2$; the \emph{elastic domain} is shown in pink; (b) A schematic structure of  the four equivalent   transformation paths  in the extended configurational space of metric tensors $\mathbf{F}$. 
}
\label{fig:9}
\end{figure}

To rationalize  the observed orientation relations, we now take advantage of the proposed  novel mapping of the results of MD simulations into  the $\mathbf{C}$-space. We begin by  showing  in Fig. \ref{fig:9}(a)  a fragment of the $\mathbf{C}$-space centered  around the  point  S  (taken in this case as the reference). It is  essentially a zoom in on Fig. \ref{fig:3}(a). 

Our  Fig. \ref{fig:9}(a)  shows more vividly that  an unbiased (pressure or temperature induced) S-T transition would be  necessarily represented  by two simultaneous and parallel transformation paths: S $\to$ T$_1$ and S $\to$ T$_2$.
 Both   describe pure shear deformations traversing configurations with rhombic symmetry. Note that,  while  in  Fig. \ref{fig:9}(a)  we show only the isochoric projection of the $\mathbf{C}$-space;  it will be explained  below that the actual  S $\to$ T$_1$ and S $\to$ T$_2$  transitions in our MD numerical experiments  also carry an attendant  volumetric contraction. 

Since the description in terms of metric tensors does not account for  rigid rotations, it cannot be used to rationalize the observed  orientation relationships between neighboring grains.  To this end we need to   advance  from
 
$\mathbf{C}$-space to the larger  $\mathbf{F}$-space,  where expectedly an even  more complex picture emerges, see our Fig. \ref{fig:9}(b). Thus, the  ``deformation variants” in $\mathbf{C}$-space multiply as  ``orientation variants” in $\mathbf{F}$-space  where the same metric can describe several  deformation states which differ by the  orientations of the  basis vectors.
 
To be more quantitative, we  observe that   along the  pure shear  path S $\to$ T$_1$   the stretch tensor    \begin{equation}
{\bf U}={\bf U}_\diamond(\lambda)=\sqrt{{\bf C}_\diamond}
   \end{equation}
    can be written in the form \cite{thiel2019shear} 
\begin{align}
{\bf U}_\diamond(\lambda)&=
\frac{1}{2}\left[\begin{array}{cc}
\lambda+\frac{1}{\lambda} & \lambda-\frac{1}{\lambda}\\
\lambda-\frac{1}{\lambda} & \lambda+\frac{1}{\lambda} 
\end{array}\right],
\label{eq:pure_shear2}
\end{align}
where  $\lambda=1$ at the square phase S and $\lambda=\lambda_*=3^{1/4}$ at the triangular phase T$_1$.  Along the apparently  synchronous path S $\to$ T$_2$ the stretch tensor is
 \begin{equation}
{\bf U}={\bf U}_\diamond(1/\lambda).
   \end{equation}

Note next, that both target mappings ${\bf U}_1={\bf U}_\diamond(\lambda_*)$ and ${\bf U}_2={\bf U}_\diamond( \lambda_*^{-1})$, correspond to area preserving stretchings along two opposite diagonals of a square lattice cell with   one of the diagonals becoming longer than the other. These mappings, however,  do not fully characterize the complete  S $\to$ T transition because the underlying rigid rotation remains unspecified.

Thus, to   ensure geometric compatibility of the variants T$_1$ and  T$_2$ with the original square phase  S , a  clockwise  rotation  $\mathbf{R}^+(\vartheta)$ and anti clockwise rotation $\mathbf{R}^-(\vartheta)$   with $\vartheta$ = $\pm$ 15$^\circ$ have to be added 
 to ${\bf U}_1$  and ${\bf U}_2$. With such rotations included, we obtain   four equivalent triangular lattices T$_1^+$, T$_1^-$, T$_2^+$ and  T$_2^-$. They are represented  schematically in Fig. \ref{fig:9}(b). The corresponding deformation gradients combining stretches with rotations can be written in the form    \begin{equation}
 \mathbf{F}_{1,2}^{\pm} =  \mathbf{R}_{1,2}^{\pm}\mathbf{U}_{1,2} ,
    \end{equation}
     where
     \begin{equation}
\mathbf{R}^{\pm}_{1}  =\frac{1}{\sqrt{\cosh\alpha}}\left[\begin{array}{cc}
\cosh(\alpha/2) & \pm\sinh(\alpha/2)\\
\mp\sinh(\alpha/2) & \cosh(\alpha/2)
\end{array}\right]
 \,,\label{eq:rhombic_path2}
\end{equation}
and  $\alpha=2\ln{\lambda_*}$. Note that   clockwise and counter clockwise rotations are denoted via `+' and `-' respectively. One can see that  the  lattice compatibility is achieved because the rotations  \eqref{eq:rhombic_path2} align  the basis vector $\mathbf{e}_1$,   which has been already rotated  by  the stretches  $\mathbf{U}_\diamond(\lambda_*) $ and $\mathbf{U}_\diamond(1/\lambda_*) $,  with the horizontal direction. 

We  have now all the elements needed to explain the observed  misorientation of the variants T$_1$ and  T$_2$ in Fig. \ref{fig:7} and Fig. \ref{fig:8}.  Let us, for instance,  evaluate  the relative rotation between  the   T$_1^+$ and T$_1^-$ neighboring  triangular grains shown in Fig. \ref{fig:9}. Recall that these coexisting  variants  are obtained from the square phase S  via   deformation gradients $\mathbf{F}_{1}^{\pm}$ = $\mathbf{R}_{1}^{\pm} \mathbf{U}_{1}$. Given that  $\lambda_*=3^{1/4}$ and  $\alpha=2\ln(\lambda_*)$, the corresponding misorientation angles are  
\begin{equation}
\vartheta_{\pm} =  \sin^{-1} \left[\pm \frac{\sinh(\alpha/2)}{\sqrt{\cosh(\alpha)}} \right] \frac{180^\circ}{\pi}= \pm 15 ^\circ 
 \,. \label{eq:theta}
\end{equation}

\begin{figure}[H]
\centering
\includegraphics[width=0.4 \textwidth]{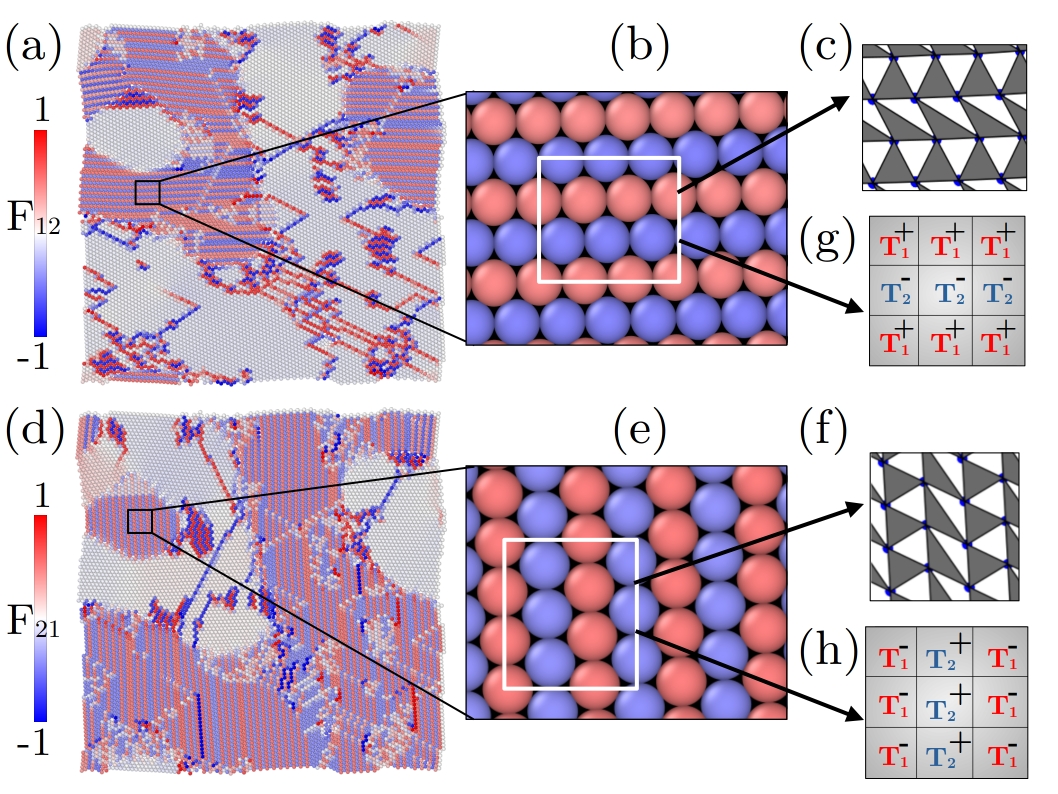}
 \caption{In (a), (d) we depict the MD simulated  fields $F_{12}$ and $F_{21}$ in the transformed triangular phase.  Zoomed-in views of fragments from (a,d) are presented in (b,e)  where we see  the corresponding  atomic configurations  visualized using OVITO \cite{stukowski2009visualization}.  In (c,f) we show triangulation representations corresponding to zoomed-in fragments from (b,e).  In (g,h) we identify the corresponding variants T$_{1,2}^{\pm}$. 
}
\label{fig:10}
\end{figure}

One can see  that the rotations  $\mathbf{R}_{1}^{+}(\vartheta)$ and $\mathbf{R}_{1}^{-}(\vartheta)$  align the basis vectors $\mathbf{e}_1$ and $\mathbf{e}_2$ of  the T$_1$ lattice  with the horizontal and vertical directions, respectively.  The resulting  misorientation angle between the  variants T$_1^+$ and T$_1^-$ is exactly  30$^\circ$ as we have  seen in Fig. \ref{fig:8}(a). Similarly, analysis for the  coexisting  variants  T$_2^+$ and T$_2^-$  shows that the corresponding misorientation angle is  again 30$^\circ$; as we show below,  the variants   T$_2^+$ and  T$_2^-$ also coexist in neighboring  grains separated by a grain boundary.   

%

Indeed, our  Fig. \ref{fig:10}(a,d)   confirms   that all four variants  T$_{1,2}^{\pm}$ are encountered in the  grain textures obtained  in the  MD numerical experiments. Moreover, one can see that both  atomic configuration, reached at the end of the S-T transformation, feature alternating rows/columns
of positive  and negative  components of the deformation gradients,  $F_{12}^{\pm} $  and $ F_{21}^{\pm} $, see Fig. \ref{fig:10}(b,e).   In other words,  we observe  two types of nano-scale mixtures:  either   T$_1^+$ and T$_2^-$, in  Fig. \ref{fig:10}(g),  or T$_1^-$ and T$_2^+$, in   Fig. \ref{fig:10}(h). In  Fig. \ref{fig:10}(c,f)  we show the associated  Delaunay triangulation   visualizing the non-affine deformation behind the apparently homogeneous lattice structure inside each of the  grains.

\begin{figure}[H]
\centering
\includegraphics[width=0.5\textwidth]{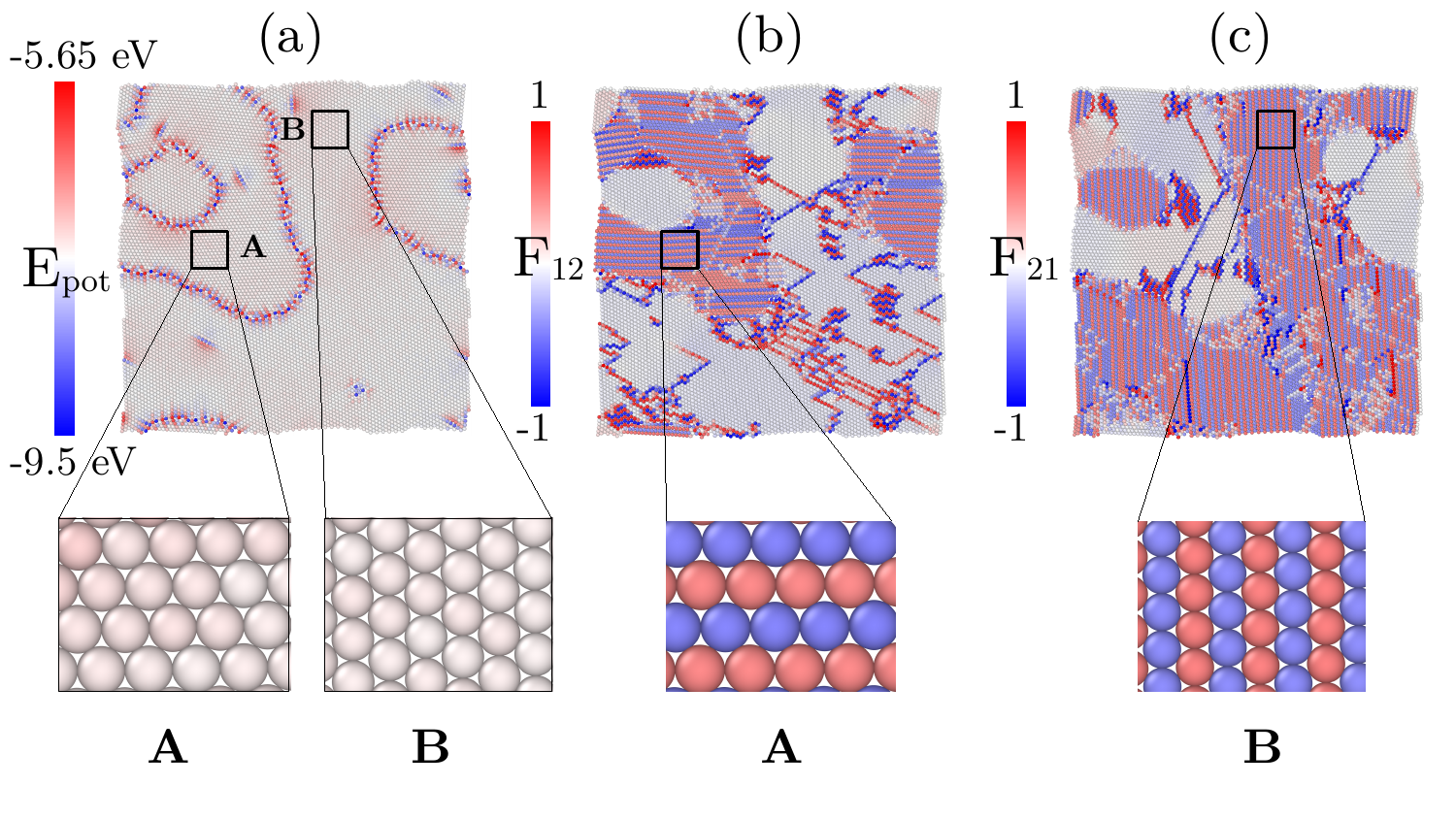}
\caption{Transformed triangular polycrystal in MD numerical experiment with atoms colored using: (a) potential energies per atom, (b) F$_{12}$ and (c) F$_{21}$. Two fragments A and B are also shown as insets.}
\label{fig:11}
\end{figure}

To  stress that  our  Fig. \ref{fig:7} and Fig. \ref{fig:10}  show the outcome of the same MD numerical experiment,  we present  them together   in Fig. \ref{fig:11}. Both figures display the same transformed triangular crystal with the only difference  that in Fig. \ref{fig:7} atoms are colored according to their potential energies, see the reproduction in Fig. \ref{fig:11}(a),  whereas in Fig. \ref{fig:10}, which we present now as Fig. \ref{fig:11} (b,c), atoms inside two different triangular phase grains  are colored  according to the value  of the particular components  of the deformation gradient:   F$_{12}$  in Fig. \ref{fig:11}(b) and F$_{21}$ in Fig. \ref{fig:11}(c).

More specifically, the  fragment A,  shown in Fig. \ref{fig:11}(a)  and reproduced  in Fig. \ref{fig:11}(b),  emphasizes  the component F$_{12}$. It presents an example of a horizontal nano-twinning

which can be also viewed as crystallographically specific finite shuffling. Instead, the  fragment B, shown  in Fig. \ref{fig:11}(a)  and reproduced  in Fig. \ref{fig:11}(c) emphasizes  the component F$_{21}$. In this case  the  nano-twinning   is  vertical. Note that the  special  coloring of the grains in Fig. \ref{fig:11}(b,c)  is chosen  in such a way that the gray areas   always indicate that either F$_{12}$=0  or F$_{21}$ = 0. 

The obtained  numerical results suggest  that  the  standard   representation  of MD data,   interpreting the outcome  of  S-T transition as a polycrystal with misoriented homogeneous grains, is deceptive. Instead, the  new way of representing such data reveals crystallographically specific  nano-twinning disguised as rigid rotations. Moreover, given that the obtained antiparallel  atomic displacements  correspond   exactly  to lattice invariant shears, it is natural to interpret the resulting pattern as representing   alternating  plastic slips.  The emerging  perspective on the nature  of the   S-T transition   complements and broadens previous studies of its mechanism
which apparently overlooked   the possibility that the product phase can be represented  at the atomic level  as a fine mixture of  elementary  variants \cite{hatch2001systematics, toledano1996recon, srinivasan2002mechanism,bhattacharya2004crystal,conti2004variational,van2019compression, damasceno2009pressure,jagla2017observation, laguna2009classical, laguna2015testing, Rechtsman2006,kryuchkov2018complex,dmitriev2023discontinuous,shuang2022atomistic}.

\begin{figure}[H]
\centering
\includegraphics[width=0.43 \textwidth]{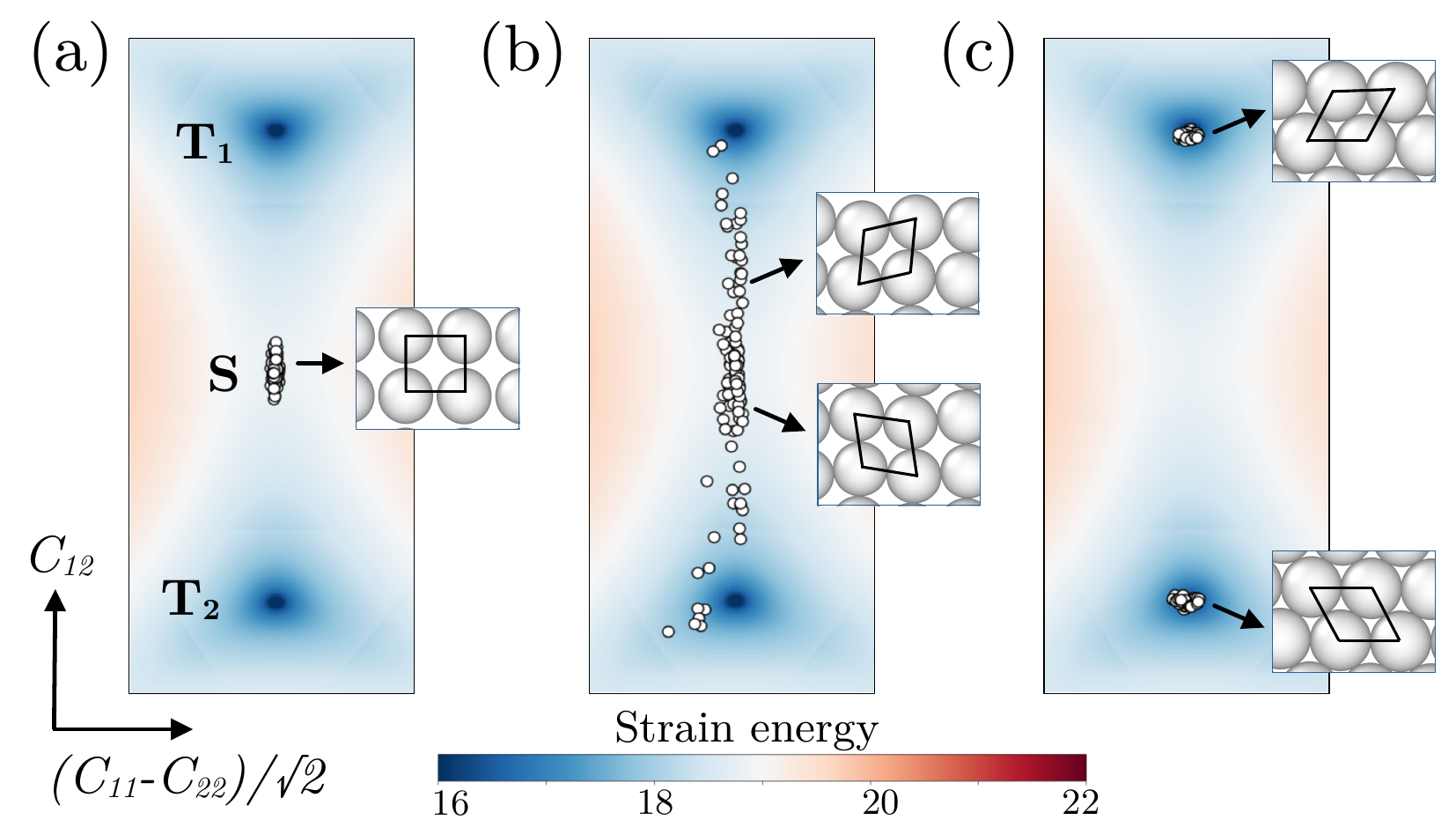}
\caption{Gradual progression of the S$\to$T transformation in $\textbf{C}$-space at three different stages  corresponding to points (1-3) indicated in Fig. \ref{fig:6}(b). Fragments of the initial square and the transformed rhombic and triangular configurations are shown in the insets. The energy landscape is visible at the background. 
} 
\label{fig:12}
\end{figure}

Note next that in our MD simulations the distribution of the values of  the metric tensors $\textbf{C}$  evolved during the S-T transition. Thus, the transformation starts when all  the values of $\textbf{C}$  were exactly  the same   and their  distribution was fully localized.  At the end of the transformation, when the T phase was  nominally reached, the configurational points spread all over the $\textbf{C}$-space.  In Fig. \ref{fig:12}, we show a fragment of the computed energy landscape  around the reference   configuration  S which  includes the   two target configurations   T$_1$  and   T$_2$. We  mapped into the same   $\mathbf{C}$-space all the atomic strains    while  showing separately the three stages (1-3) of the S-T transition indicated in Fig. \ref{fig:6}(b).

Specifically, in Fig. \ref{fig:12}(a) we illustrate  the very beginning of the transformation when all atomic strains populate the marginally stable square  configuration  S located at the origin. At the intermediate stage of the transformation, shown in Fig. \ref{fig:12}(b), we observe spreading of 

\noindent atomic strains along the two symmetric rhombic (pure shear) paths. The configurational points visibly advance   towards two energy wells representing the variants   T$_1$ and T$_2$. We stress that both  paths are pursued simultaneously.  As a result, the transformed triangular phase emerges as  comprised  of strains populating  almost equally both  target energy wells T$_1$ and T$_2$, see Fig. \ref{fig:12}(c).






 
\begin{figure}[H]
\centering
\includegraphics[width=0.45\textwidth]{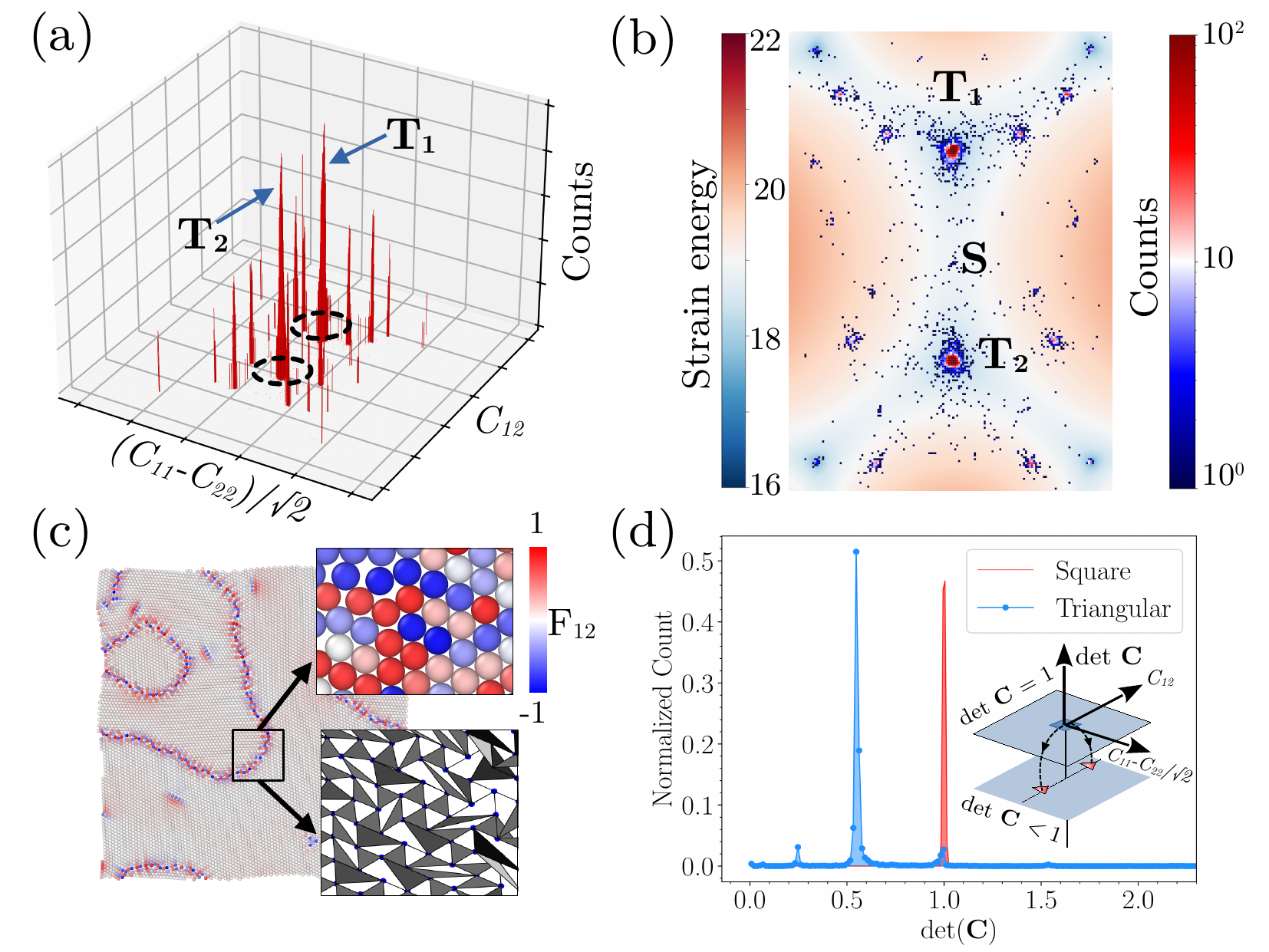}
\caption{The S-T transition  in the $\mathbf{C}$-space. (a) 3D histogram representation of the strain distribution in triangular phase;   `counts' axis has a  logarithmic scale. 
 (b)  the same strain distributions shown against    the corresponding energy landscape. (c) A highly deformed atomic fragment around the grain boundary;    both deformation gradient distribution and the  deformed triangulation network are shown in the two insets. (d) The distribution of det$\mathbf{C}$ in the initial square (red) and the final   triangular lattices (blue).  
 }
\label{fig:13}
\end{figure}

Our  Fig. \ref{fig:13} illustrates  in more detail   the final strain distribution.  The   histogram representation of the strain distribution in Fig. \ref{fig:13}(a)   shows in logarithmic scale  that most of the elements are  in either T$_1$ or  T$_2$ energy wells; this is also seen in its 2D projection shown in the Fig. \ref{fig:13}(b). 

\begin{figure}[H]
\centering
\includegraphics[width=0.48 \textwidth]{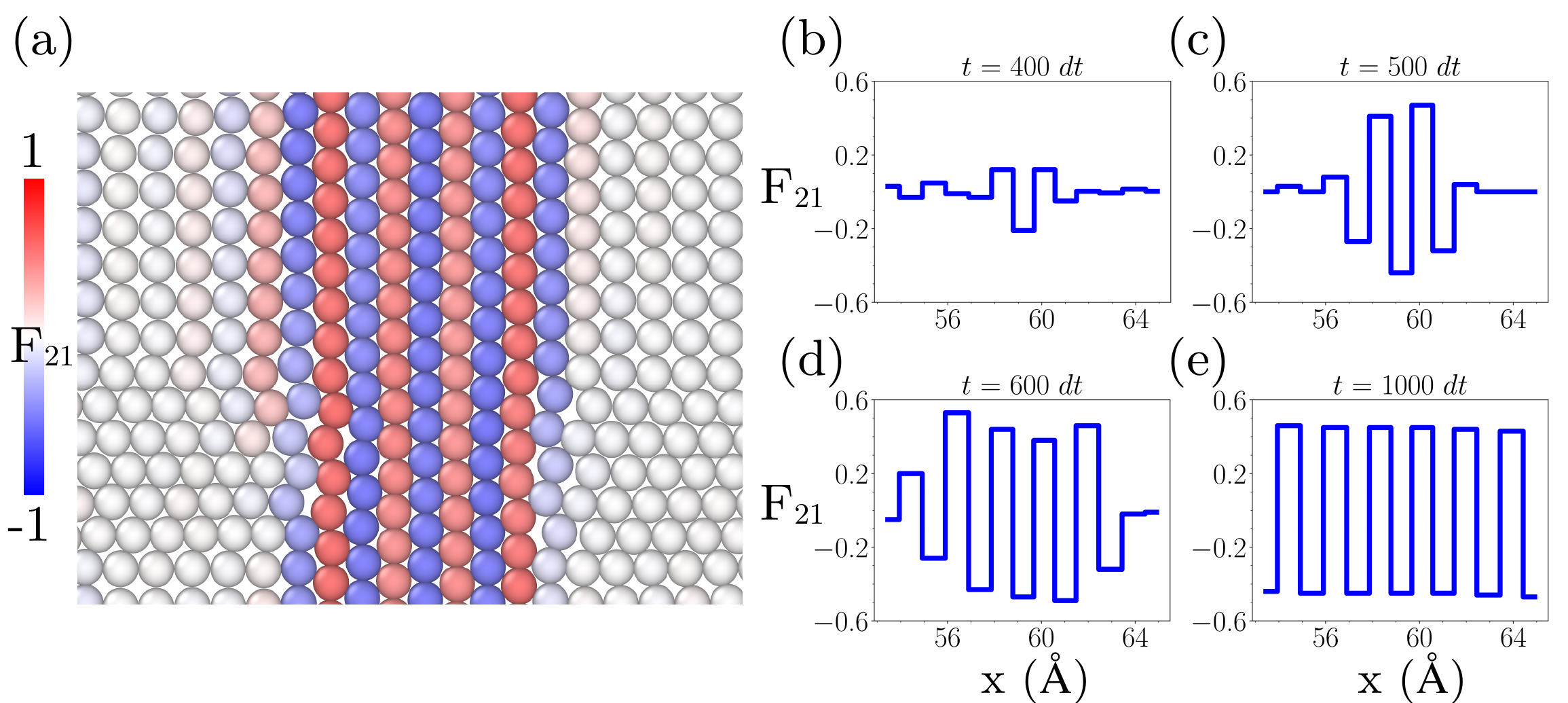}
\caption{(a) Layer-by-layer propagation of the deformation gradient (here only $F_{21}$ component)   during S-T transformation;
(b-e)  transient spatial configurations of the deformation gradient showing the formation of an ideally periodic pattern.}
\label{fig:14}
\end{figure}

Note however, that in addition to the  most populated triangular configurations T$_1$ and T$_2$, several other locations  outside the  T$_1$ and T$_2$ energy wells  are also  occupied. Most of them  reflect the structure of  grain boundaries  like the one  shown in Fig. \ref{fig:13}(c). In  Fig. \ref{fig:13}(d) we illustrate the fact that the S-T  transformation is accompanied by a volumetric contraction as  the original single crystal S phase with det$\mathbf{C}$ = 1 finally transforms into the final polycrystal T phase with  det $\mathbf{C}$ = 0.55.




A more detailed  analysis of the stage-by-stage  transformation process in the physical space, illustrated  in Fig. \ref{fig:14}(a), shows how the alternating micro-slips,  represented by interdigitated fields $F_{12}$  or $F_{21}$, are actually  developing. One can see that the apparently shuffled microstructure grows layer-by-layer. More specifically, the associated nano-scale `zipping' takes place  through back and forth transverse propagation of Shockley partials. The possibility of such  coordinated  motion of surface steps has been also observed in other systems   \cite{bejaud2018effect}. At macro scale this micro dynamics remains hidden and  the transformation   appears as proceeding through front propagation.  While such front leaves behind a pattern of anti-parallel micro-displacements,  what emerges at the macroscale is a rigid rotation of a perfect triangular lattice, see Fig. \ref{fig:14}(b--e).

We stress that  the revealed microscale pattern remains concealed  behind   the conventional interpretation of  MD experiments which would  present the product phase as homogeneous inside each of the grains.  Instead, the proposed novel way of interpreting MD data shows that  such  apparently homogeneous phase  is  a disguised  atomic scale mixture of two different but equivalent  energy  wells.  Since the corresponding variants   are geometrically compatible and the interfaces between them are energy free, the effectively  plastic   nature of such  apparent  lattice  rotations  has been so far hidden.

To complement the obtained picture of the direct S-T transition,  we have also performed some numerical experiments where we also observed   the reverse T-S transition.  More specifically, we performed MD cyclic loading of our samples. The resulting hysteresis loop can be viewed as representing a succession of two,  S$\to$T  and  T$\to$S,   reconstructive transitions. 

\begin{figure}[H]
\centering
\includegraphics[width=0.45\textwidth]{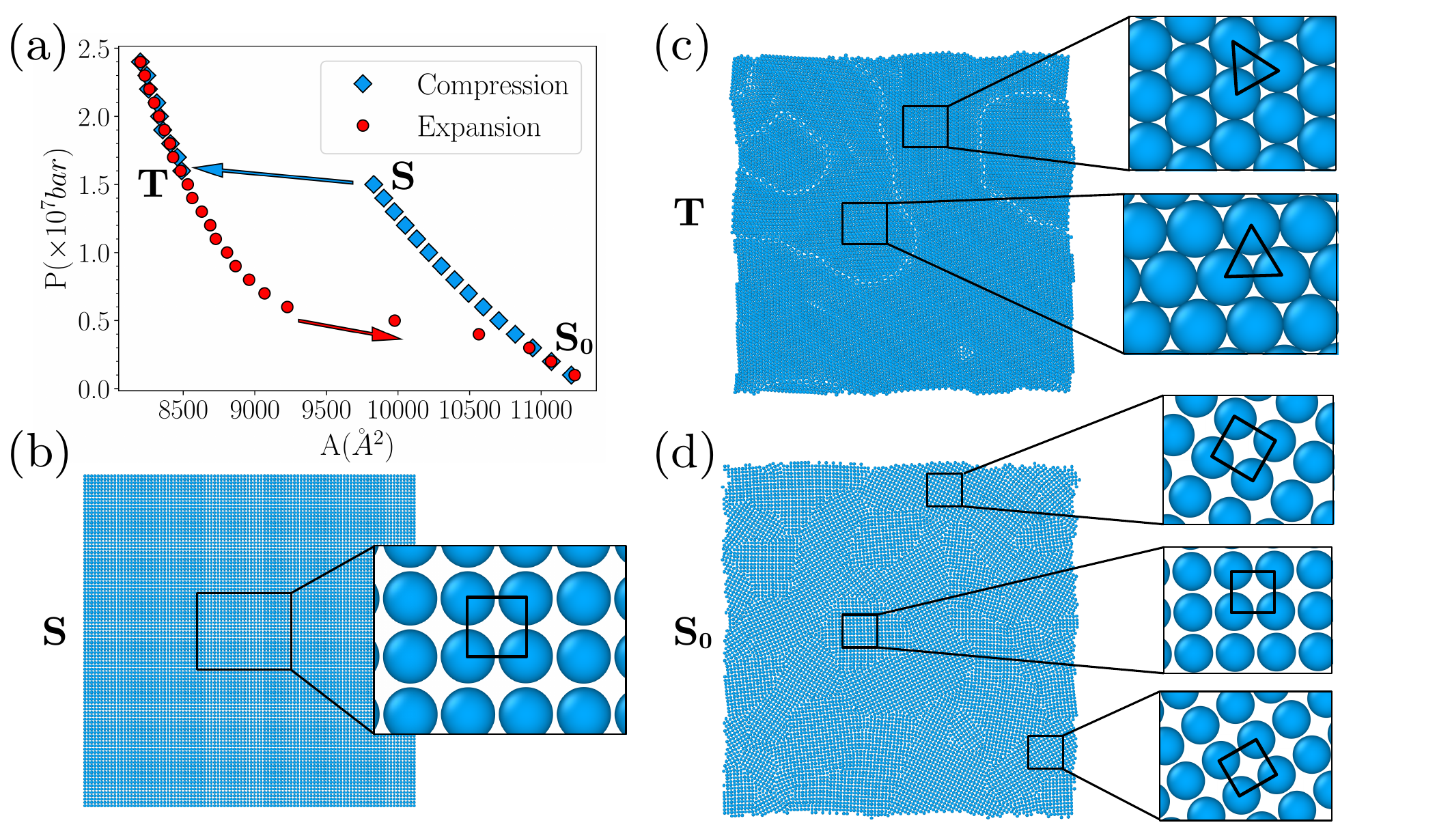}
\caption{S$\to$T$\to$S  reconstructive transitions in preliminary MD simulations.  (a) P-V(area A) phase diagram of the complete compression-expansion cycle. With the letters S, T, and S$_0$ we denote :  pristine square, transformed triangular lattice under compression and transformed square lattice after the subsequent expansion. Insets in (b), (c), and (d)   show microscopic  configurations  with their associated multi-grain compositions.}
\label{fig:15}
\end{figure}

In Fig. \ref{fig:15}(a) we  show  the implied  compression-tension sequence on the  P-A  plane.  As in our previous  numerical experiments, we started with a defect-free (pristine)  2D square crystal, see Fig. \ref{fig:15}(b),  and performed isotropic compression (by increasing hydrostatic pressure) to induce the  transformation into a triangular phase. The resulting configuration of misoriented triangular grains at P = 1.6 $\times$ 10$^7$ bar is shown in Fig. \ref{fig:15}(c). We then further slightly increased the pressure up to 2.4 $\times$ 10$^7$ bar before reversing the direction of loading through  gradual   expansion  the system  via the reduction of pressure. This brought  the system back into the square phase which was no longer homogeneous. Instead,  we  observed  a texture of misoriented square grains shown in Fig. \ref{fig:15}(d). While the  nature of the underlying rotations will be discussed separately, here we only mention that successive S$\to$T$\to$S$\to$T$\to$ ... transformations progressively increase the complexity of the variant mixture with eventual convergence to a  plasticity-dominated shakedown state.

\section{Molecular statics} \label{sec:athermal_results}

In addition to standard finite temperature MD numerical experiments,  we also used athermal  zero temperature  molecular statics (MS)  protocols.  This allowed us to investigate the sensitivity of the observed features of the S-T transition to temperature. 

\begin{figure}[H]
\centering
\includegraphics[width=0.45\textwidth]{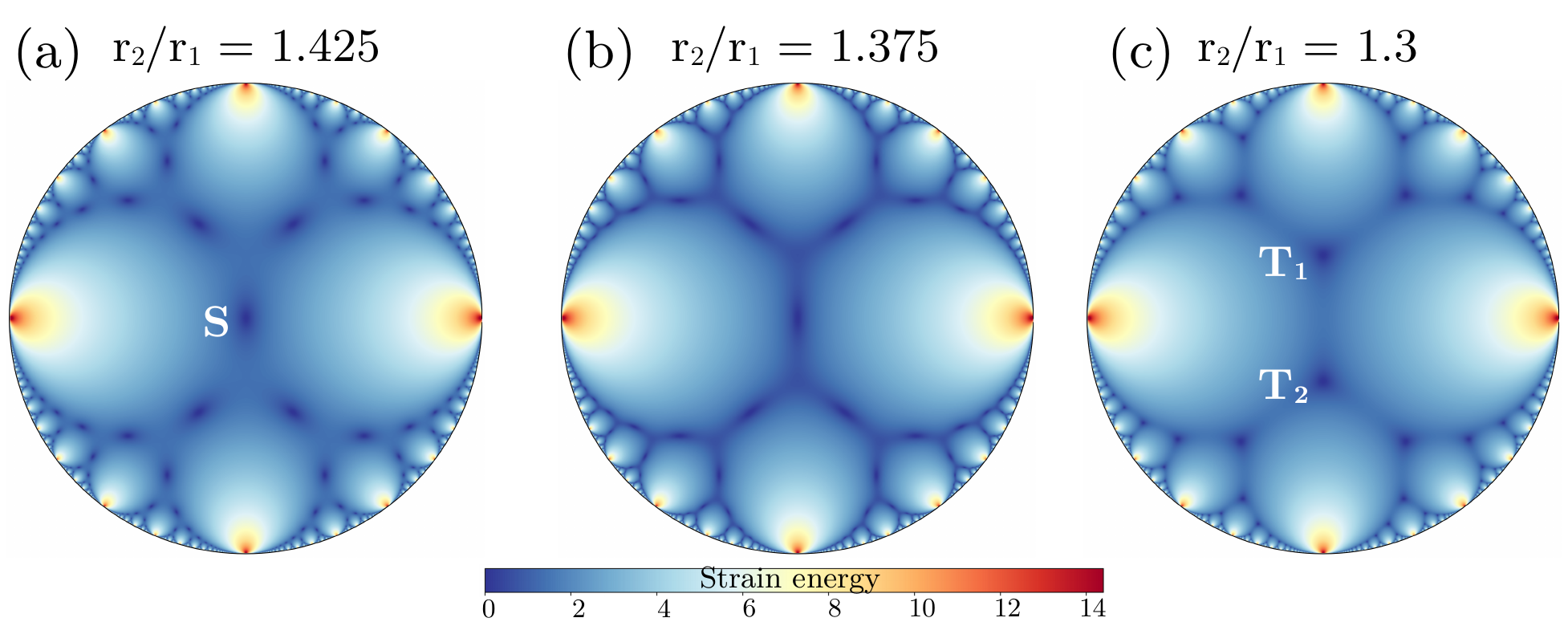}
\caption{Strain energy landscape   (color coded) in the  configurational $\mathbf{C}$-space (Poincaré disk) at three different values of the Boyer parameter r$_2$/r$_1$. Letters indicate the square (S) and the two relevant triangular   minima (T$_1$, T$_2$). }
\label{fig:16}
\end{figure}

Recall that in  MD simulations, we used the external  pressure  to induce the transition. Since in MS   the  thermodynamic pressure is an ambiguous concept  and only virial pressure can be computed confidently, we have chosen  to induce the S-T transition by changing the potential parameter of the  potential \cite{boyer1996}. Specifically, to observe     triangular lattice we   lowered the parameter  r$_2$/r$_1$ from the original value  1.425 to 1.375 where   the square lattice was almost elastically marginalized.  In Figure \ref{fig:16} we 
 
%

illustrate  the corresponding evolution of the energy landscape in the $\mathbf{C}$-space (Poincaré disk). One can see that     the  location  of   energy minima  shift from square configuration  `S' in Fig. \ref{fig:16}(a)  to triangular configurations  `T$_1$' and `T$_2$' in Fig. \ref{fig:16}(c).

Periodic boundary conditions were maintained throughout the simulation. Initially, a stable planar square crystal was prepared with $10^4$ atoms   using the same potential as in our MD simulations.  Then, to induce the S-T transition in a near marginal state at  r$_2$/r$_1\approx1.375$, we introduced  a small disturbance  by displacing  all the atoms  at random distances (about  $0.9$ $\%$ of the lattice parameter)  along both $x$ and $y$ directions. Afterwards, the parameter was lowered further till the value  r$_2$/r$_1=1.3$ where the  instability took place and the conjugate gradient  algorithm was used to perform local energy minimization and to locate  the new equilibrium configuration.

\begin{figure}[H]
\centering
\includegraphics[width=0.5\textwidth]{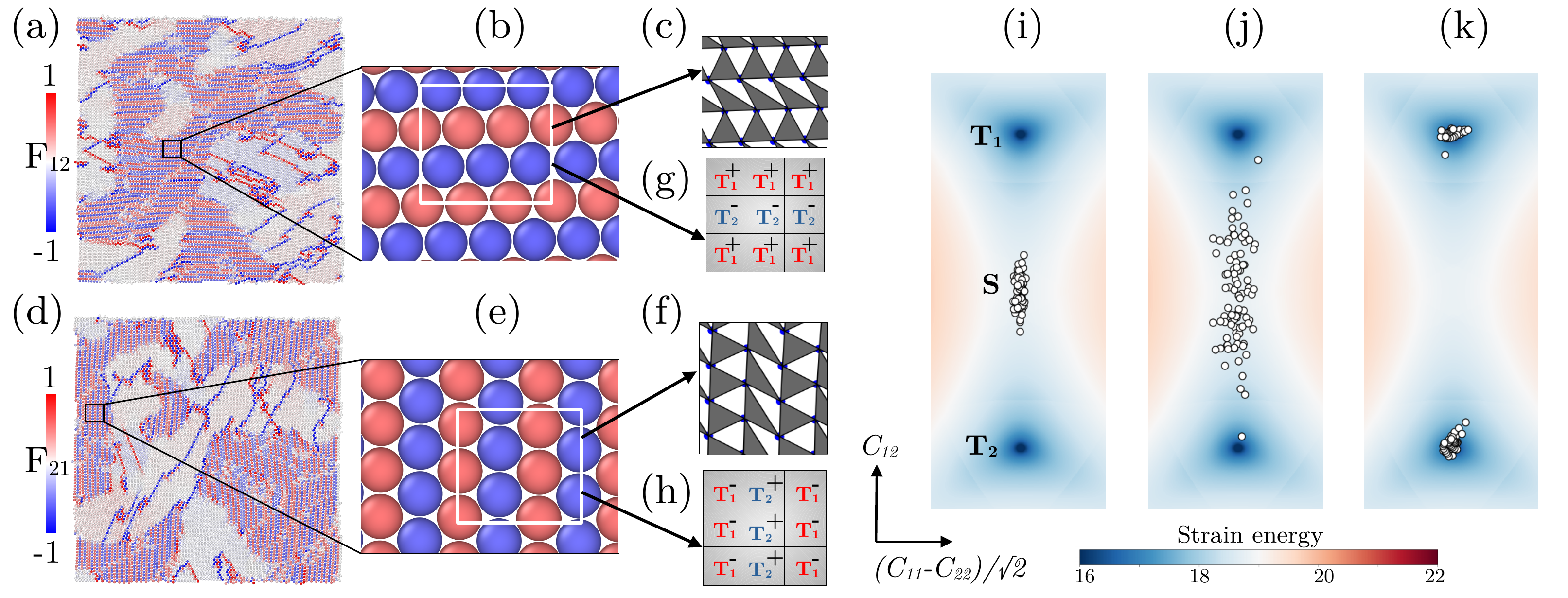}
\caption{(a), (d) The distribution of deformation gradient components $F_{12}$ and $F_{21}$ in the transformed triangular phase obtained in MS simulations. Zoomed-in views of
fragments from (a,d) are presented in (b,e)  and triangulation representations of (b,e) are shown in (c,f) respectively. (g,h) shows the corresponding variants T$_{1,2}^{\pm}$.  (i)--(k) Evolution of the atomistic strain distribution in the $\mathbf{C}$-space.}
\label{fig:17}
\end{figure}

Our numerical simulations  of  S-T transition using  MS protocol  exhibited  all the main   elements  of  the   transformation mechanism observed in MD experiments.  In particular, our Fig. \ref{fig:17}(a--h) show   that the lattice scale alternate plastic slips involving both, atomic rows and  atomic  columns, have been  recovered.  More specifically,  we observed in neighboring grains the same  alternating mixtures of configurations T$_1^+$, T$_2^-$ (realized via alternating $F_{12}^{\pm}$) and of configurations  T$_1^-$, T$_2^+$ (realized via

alternating $F_{21}^{\pm}$),  see Fig. \ref{fig:17}(g,h). The evolution of the strain populations inside the $\mathbf{C}$-space   indicates basically the same  mechanism involving concurrent  symmetric pure shears, see our Fig. \ref{fig:17}(i--k).  As in our MD experiments,  the  spreading of atomic strains via rhombic valleys towards  the triangular energy minima T$_1$ and T$_2$ took place in the form of propagating fronts separating the receding micro-homogeneous state from the expanding  micro-inhomogeneous, pseudo-shuffled mixture states.

\begin{figure}[H]
    \centering
    \includegraphics[width=0.43\textwidth]{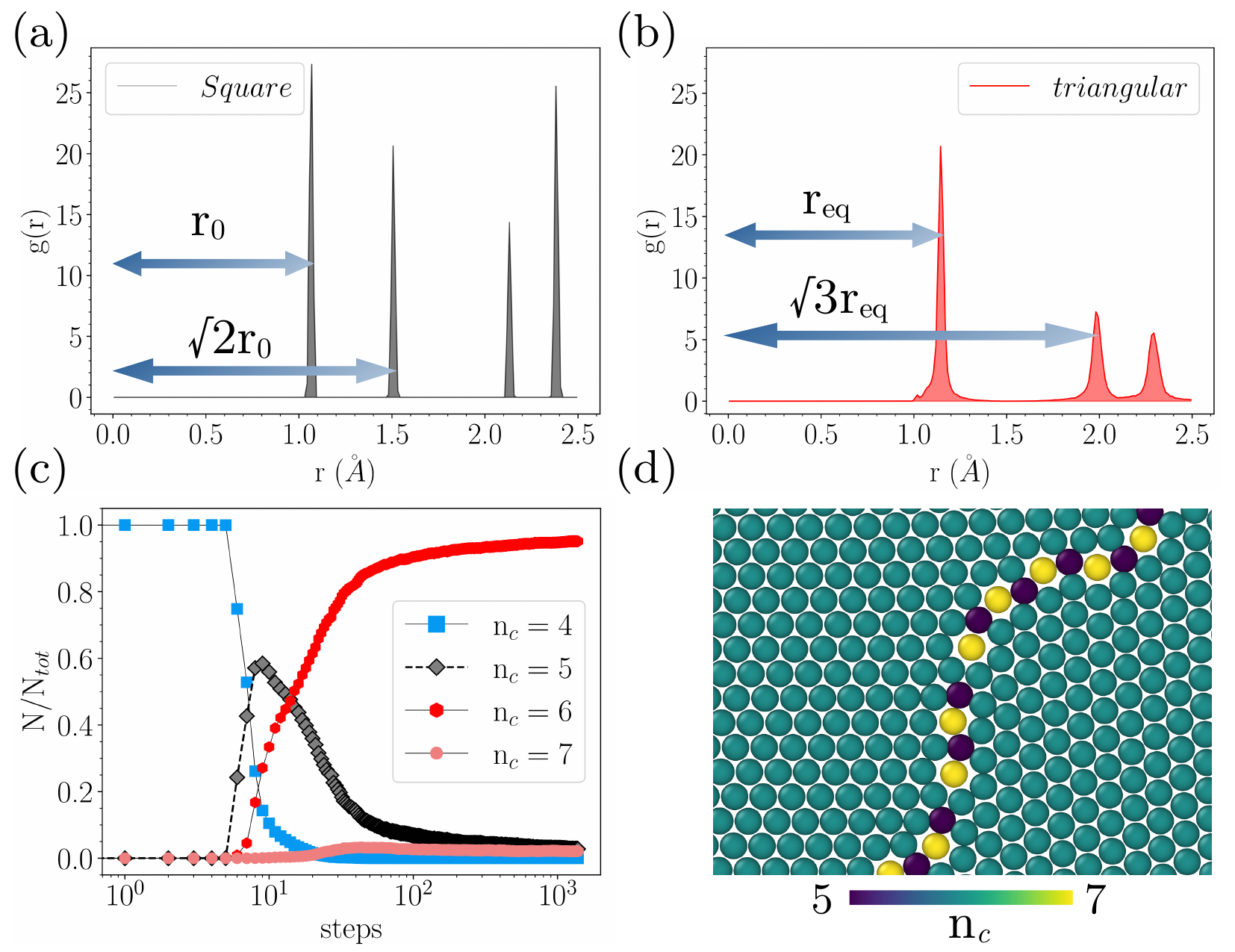}
    \caption{Radial distribution functions $g(r)$  for the initial    square lattice (a) and for the  transformed triangular  crystal (b). In (c) we show the variation during the  S-T transition  of the fractions of atoms (N/N$_{tot}$) characterized by  different coordination (n$_c$).  A fragment of the misoriented triangular grains with atoms colored according to their  coordination is shown in (d).}
    \label{fig:18}
\end{figure}

A minor  difference between MD and   MS   numerical experiments is that in the latter   the S-T transformation proceeded in almost isochoric conditions.  To explain this effect we computed the  radial distribution function   
   \begin{equation}
   g(r) = \frac{1}{\pi r^{2}N\rho} \sum_{i=1}^{N} \sum_{j\neq i}^{N} \left\langle \delta (\bf{r} - |\textbf{r}_\textit{j}-\bf{r}_\textit{i}|) \right\rangle,
      \end{equation}
  where  $\textbf{r}_{j}$ - $\textbf{r}_{i}$ is  the distance between the atoms $i$ and $j$, $r=|\bf{r}|$,   $\rho$ = $N/A$ is the density, $N$ is the total number of atoms,  $A$ is the area of the system and the averaging is over angular variables.   As we show in Fig. \ref{fig:18}(a,b), the 
 S-T transition results in the change of the value of the lattice constant  from 1.0659 \AA{} [r$_0$ in Fig. \ref{fig:18}(a)] in the `S' phase to 1.14 \AA{} [r$_{eq}$ in Fig. \ref{fig:18}(b)]
in the `T' phase. The corresponding areas of the unit cell are: r$_0^2$ = 1.1361 \AA$^2${} for the square  lattice and   $\frac{\sqrt{3}}{2}$ $r_{eq}^2$ = 1.125 \AA$^2${}   for the  triangular lattice  (with its rhombic cell).  While the ratio of the two areas is almost equal to one,   $\frac{  area_{\square}}{area_{\triangle}}$ = 1.009, still  during   S-T transition the packing fraction  increases. Indeed, if the  triangular/hexagonal lattice is closed-packed,  its square counterpart is not. Specifically, the  packing fraction, defined as the ratio of the area occupied by the atoms inside the unit cell and that of the

\begin{figure}[H]
    \centering
    \includegraphics[width=0.42  \textwidth]{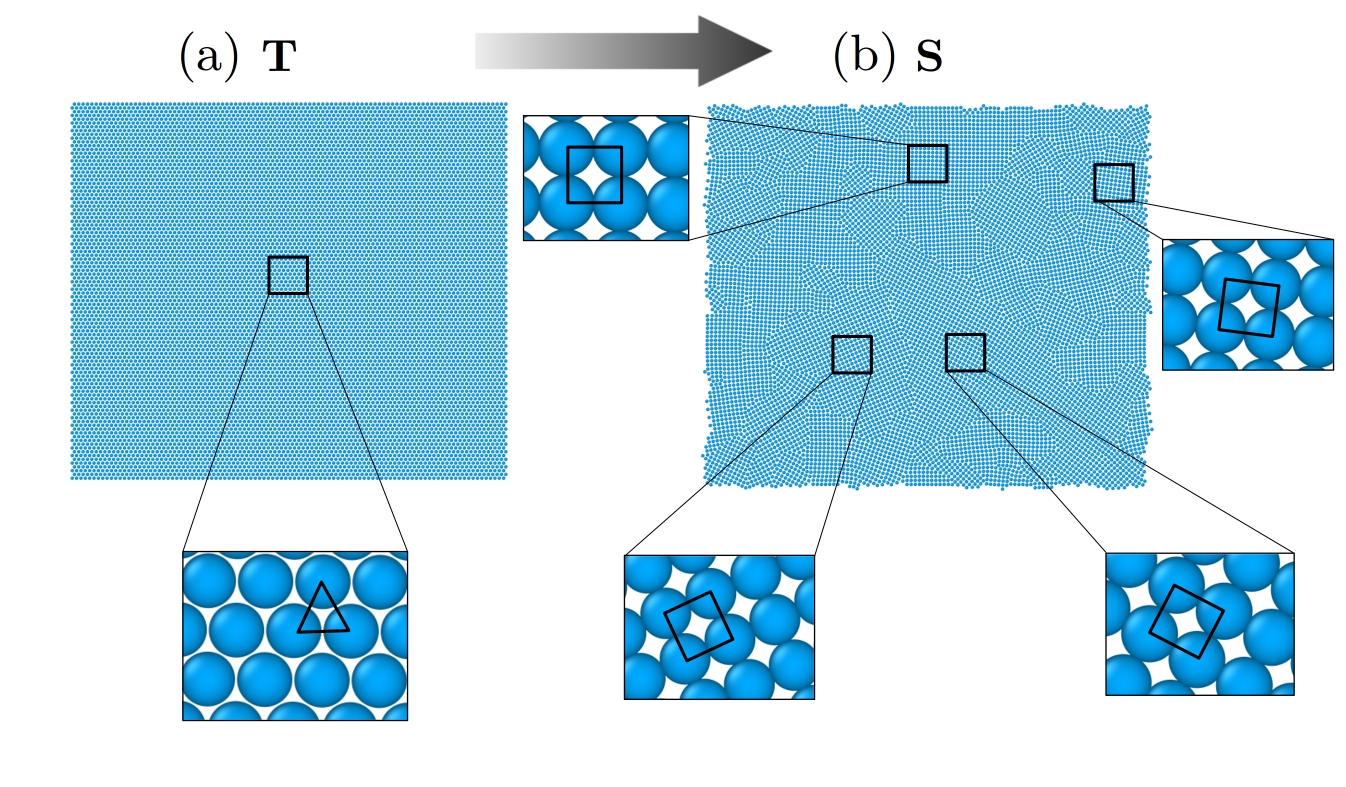}
    \caption{Preliminary modeling of T$\to$S transition using athermal MS simulations: (a) pristine triangular lattice (T), (b) transformed square (S) lattice  represented by a texture of misoriented grains. Fragments of `T'  and `S' grains with their associated  orientations are   shown in the  insets.}
    \label{fig:19}
\end{figure}

area of the unit cell,  is $\frac{\pi a^2}{4}/a^2$ = $\frac{\pi}{4}$ $\approx$ 0.785 in `S' phase while it is $\frac{1}{2}\pi\frac{a^2}{4}/\frac{\sqrt{3}a^2}{4}$ = $\frac{\pi}{2\sqrt{3}}$ $\approx$ 0.907, in  `T' phase   see the  distribution of the  coordination  numbers   n$_c$ in Figure \ref{fig:18}(c).   Therefore, in view of the similarities of the parameters of the unit cells,  the zero  volumetric  effect  in such the S-T transition can only  result from the emergence   of  `void space' that is realized  through the formation of less dense  grain boundaries, see  our  Fig. \ref{fig:18}(d). In particular, we see that the atoms forming the   grain boundaries   are differently coordinated with n$_c$ = 5,7  vis-à-vis the atoms inside the  triangular grains where n$_c$ = 6. More specifically, our numerical experiments showed  that around 5 $\%$ of atoms contributed towards  loose  grain boundaries  against 95  $\%$ of atoms inside the close packed grains. None of these purely geometrical zero temperature arguments is, of course,  applicable in the case of finite temperature MD simulations. 


Finally, to corroborate the  results obtained in our MD experiments regarding   the reverse T-S transition,   we  also performed  the   athermal MS simulation of the corresponding  unloading tests. Specifically,  we simulated   the transition from a pristine triangular phase  to a multi-grain  square phase by increasing the value of the parameter  r$_2$/r$_1$ from 1.3 to 1.425.  As we have already mentioned, this  modifies the ground state of the system shifting the preference from triangular to  square lattice as a ground state. The resulting grain texture, see   our Fig. \ref{fig:19},    exhibits the same misorientation angles as in our MD simulation and also  reveals a hidden alternating slip distribution behind the apparent homogeneous rotations of the product phase. A systematic study of the cyclic S-T-S transition  is underway and the discussion of the detailed structure of the resulting plastified lattice configurations is left for a separate study.


\section{Mesoscopic tensorial model}

While both MD and MS based  numerical experiments  provide   fully detailed description of transformation-induced atomic rearrangements in our model crystal  and  therefore  accurately represent micro-mechanism  of the S-T phase transition  while relying minimally on phenomenology,  such   approaches are prohibitively computationally expensive when one turns to  fine details of emerging multi-scale microstructures requiring  consideration of larger systems. Also, as we have seen,  the problem of adequate mapping of  MD and MS results on the macroscopic description of mechanical response in terms of such macroscopic observables as stresses and strains, is not yet fully resolved. A reasonable  conceptual trade-off between continuum  and atomic descriptions can be achieved using  the Landau-inspired  mesoscopic  coarse grained analog of MS  introduced in \cite{ericksen2008on,conti2004variational,salman2011minimal,baggio2019landau}. It resolves in (quasi) continuum setting the full crystallographic symmetry, including lattice-invariant shears while  accounting geometrically adequately for both large strains and large rotations \cite{salman2012critical,zhang2020variety,baggio2023homogeneous,baggio2023inelastic,perchikov2024quantized}. In what follows, we refer to this hybrid discrete-continuum computational approach as the mesoscopic tensorial model (MTM).

 To answer the question whether the MTM approach  captures the  main elements of the revealed  mechanism of S-T transformation,  we used the same globally  periodic  Landau potential $\phi(\mathbf{C})$  which was constructed above using the Cauchy-Born rule. To facilitate comparison, we used   the same model interatomic potential \cite{boyer1996} as in our MD and MS numerical experiments.
 
%

To produce a mesoscopic description of a crystal the MTM approach postulates    that the  potential  $\phi(\mathbf{C})$  describes  mechanical response of  elastic finite elements whose size is viewed as a mesoscopic cut-off spatial scale.   The piece-wise affine deformation of the  elements is presented in the form ${ \mathbf{y}} ({ \mathbf{x}})={\mathbf{y}}_{ij} N_{ij} ({\mathbf{x}})$, where ${\mathbf{y}}_{ij} $ is the deformation of the 2D network of discrete nodes  and   $N_{ij}({\mathbf{x}})$ are  linear shape functions. The elastic energy $\phi(\mathbf{C})$   associated with  the node $\mathbf{x}$ is computed under the assumption that   $\mathbf{C} = \boldsymbol{\nabla} \mathbf{y}^{T} \boldsymbol{\nabla} \mathbf{y}$.  The use of piecewise linear deformation field  $\mathbf{y}(\mathbf{x})$   turns the equilibrium problem into finite dimensional   parametric minimization of the energy functional
 \begin{equation} \label{W2}
W = \int_{\Omega_0}  \phi (\mathbf{C}) d^2x
   \end{equation}
 where $\Omega_0$ is the computational domain. 
 
 In our numerical experiments, the finite dimensional minimization  of \eqref{W2}  was accomplished using a variant of conjugate gradient optimization known as the L-BFGS algorithm \cite{bfgs2013}. 
 This algorithm seeks solutions to the equilibrium equations  
   \begin{equation}
 \partial W / \partial \mathbf{u}_{ij} = \int_{\Omega_{0}} \mathbf{P} \nabla \mathcal{N}_{ij} d^2x = 0,
    \end{equation}
where   
 \begin{equation}
\mathbf{P} = \frac{\partial \phi }{\partial \boldsymbol{\nabla} \mathbf{y}},
   \end{equation}
and  $\mathbf{u}_{ij}$ denote the values of displacement at node $ij$. The equilibrium solutions reachable through this 

\begin{figure}[H]
\centering
\includegraphics[width=.5 \textwidth]{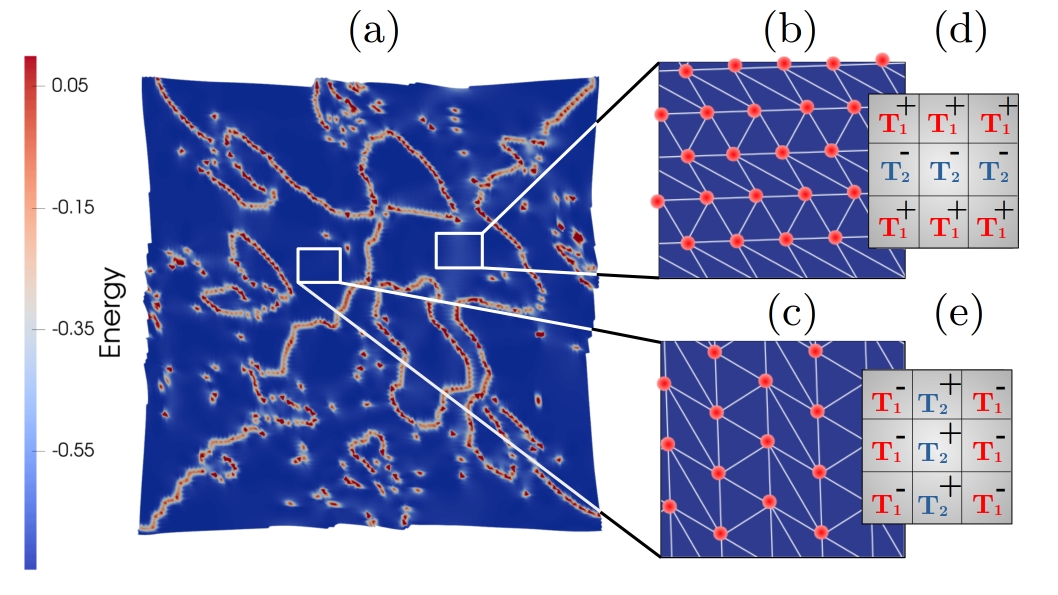}
\caption{Outcome of the coarse grained modeling of  S-T transition: (a) The post-transformation   polycrystalline texture; (b-c)   two fragments of   mis-oriented  triangular grains, with (b) presenting   (T$_1^+$,T$_2^-$) variant mixture and (c) representing (T$_1^-$,T$_2^+$) variant mixture.}
\label{fig:20}
\end{figure}

algorithm are then determined by algorithmically defined, effectively overdamped dynamics.

%

In our numerical experiments we used (discretized)  free boundary conditions  $\mathbf{P} \cdot \mathbf{N} = 0$, where $\mathbf{N}$ is the normal to the surface at the reference state. The S-T transformation was again initiated by incrementally changing the potential parameter r$_2$/r$_1$ from the value 1.425 to the value 1.3.  As we have already seen, this ensures the   shift in the nature  of the ground state configuration from square to triangular.

%
%


%
%

%
%

The results of our MTM-based  mesoscopic modeling of the S-T transformation are summarized in Fig. \ref{fig:20}. As in our molecular simulations we started with a perfect square lattice  and then brought  to a marginally stable state.  In the emerging polycrystalline configuration, illustrated  in  Fig. \ref{fig:7}, we  again observe  a   texture of  grains representing triangular lattices showing   average mis-orientation of about 30$^\circ$. The boundaries of the grains are   dislocation-rich even if here  the dislocation cores  are blurred  at the cut-off scale.

\begin{figure}[H]
\centering
\includegraphics[width=0.45\textwidth]{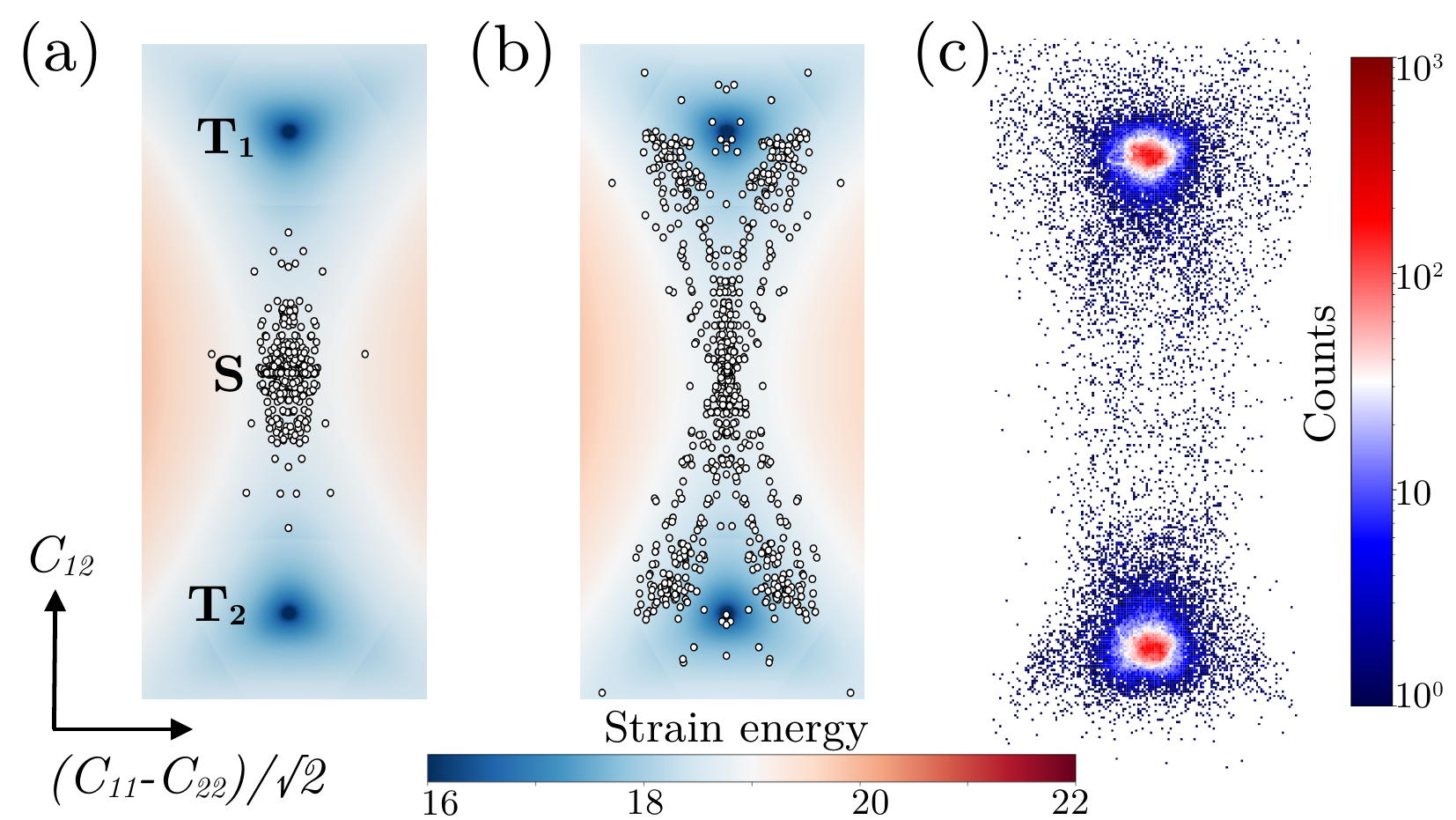}
\caption{S-T transformation in  MTM numerical experiments: (a,b)   fragments of the $\mathbf{C}$-space illustrating two successive stages of S-T transition; (c) the histogram showing the distribution of the values of $\mathbf{C}$  in the final state.}
\label{fig:21}
\end{figure}

Our Fig. \ref{fig:21} shows that the transformation is again advancing along two concurrent rhombic (pure shear)

paths, reproducing qualitatively the mechanism which  we have already seen  operating in  MD and MS simulations.   In particular, in the final configuration,  which can be viewed as  an atomic-scale mixture  of the  states  T$_1$ and T$_2$,  the  relative  rotations are again achieved through alternating crystallographically specific slips inside the   adjacent  planes  which implies  the formation of the   variant mixtures (T$_1^+$,T$_2^-$) and  (T$_1^-$,T$_2^+$), see Fig. \ref{fig:20}.  Also, similar to what we have seen in MD and MS, here we see that  transformation advances  by layer-wise `zipping'  (both horizontal and vertical) of  alternating  triangular variants.    
  
 These obtained results suggest that even within a coarse-grained description, the essential ingredient of the revealed plasticity-centered transformation mechanism can be retrieved indicating that this feature of the S-T reconstructive transition is robust.  In particular,  within the coarse-grained description  the  front-propagation-based mechanism of S-T reconstructive transition  is preserved producing  inelastic rotations of  triangular lattices  which are, in essence, micro-plastically deformed.


%

\section{Discussion}

In this paper  we showed that tracking  the history of atomic-scale  metric tensors in MD simulations offers a unique perspective on the intricate micro-pattern formation during reconstructive phase transitions. The proposed interpretation of MD numerical experiments  reveals  previously hidden details of  the deformation paths  allowing one to  analyze systematically the underlying relation between elastic and inelastic  modes. 
 
It is   appropriate to discuss  here our results  vis-à-vis the  closely related  previous 2D modeling  work of Kastner  \textit{et al.}  \cite{kastner2003molecular, kastner2006molecular, kastner2009mesoscale, kastner2011molecular} who obtained somewhat  similar  conclusions  despite considering a  conceptually different model. In their work Kastner \textit{et al.}  used binary Lennard-Jones potential to model S-T transition in a   2D  double-lattice where   sub lattices were displaced by a shift/shuffle which served in their model as an independent order parameter. During their version of S-T transformation nested square unit cells were sheared into diamonds and then  the interstitial atoms where first shuffled towards one of the sharp-angled corners, producing the hexagonal structure which then incorporated  both  sub-lattices.

Note first that in such a model there are two shear- and two shuffle-directions possible, thus \emph{four} variants of hexagonal (triangular) phase can be identified. Instead, in our model of S-T transition in monoatomic lattices there are no shuffles and therefore  there are only \emph{two} variants of hexagonal phase. 
In fact, the main  message of our work is that  shuffle can be understood as nano-twinning involving only  two variants (two energy wells). In other words, we show that quasi-shuffling does not need to be postulated separately: it can emerge in a model with two martensitic variants (instead of four)   in a form of alternating  nano-twinning which is a fundamentally novel observation. More generally,  Kastner   \textit{et al.} interpreted  their   MD  simulations as a model of  `weak' Landau-type martensitic phase transitions in shape memory alloys  where the role of plasticity is usually minimal   as their results  also confirmed. Instead, our model  deals with `strong'   reconstructive  phase transitions where plasticity is usually thought to be playing a crucial role  as it is also convincingly demonstrated by our work. We reiterate that  the problematic interpretation  of S-T transition as `weak' is due to the fact that the two wells T$_1$ and T$_2$ are located exactly on the boundary of the \emph{same} fundamental domain if the latter is centered around a square phase. However, as we show, they clearly belong to two \emph{different} elastic periodicity domains if we center them around a triangular phase. In this perspective, mixing of these two wells should be considered as plastic rather than elastic deformation.

Despite the different modeling assumptions,  Kastner \textit{et al.} also observed  in their numerical experiments the emergence of compatible  twin variants with no lattice misfit and effectively zero interfacial energy.   Their   twins  are characterized by alternating shear directions of unit cells (but with identical shuffle directions of sub-lattices)  forming a ``herring-bone pattern." However, in contrast to our observations, their twinning  mostly  takes place at meso or  macro scale. In particular, since they did not have our method of history recovery, they could not see whether  their rotated grains are internally nano-twinned. Still, the model of  Kastner \textit{et al.} apparently allowed for the formation of some irreversible `defects'.  Thus, they showed that during  the reverse T-S transformation (unloading), their model material exhibited some  plastic slip  producing point defects  which either glided to the surface, forming a kink, or piled up at obstacles in the bulk.  In our terms, they observed a nano-scale mixing of the variants of the product phase which, however,  only took place in the form of  isolated defects.  Instead, we observed that   nominally plastic deformation takes place as  a bulk phenomenon, in particular, it is responsible for the relative rotation of crystalline grains. In fact, we anticipate that the micro-mechanisms which we showed to be  operative during our prototypical S-T transitions,  contain  some generic elements   common to   most  reconstructive transitions  including the iconic  BCC-HCP and FCC-HCP  transitions.

To draw  some   specific parallels between the observed transformation paths  in 2D  and  the   mechanism of, say,  a reconstructive  BCC-HCP transition in 3D \cite{dewaele2015mechanism,gao2020twinning, bhattacharya2004crystal}, we first observe  
that  the latter involves volume preserving  pure shear deformation  in addition to  shuffling.  However,  these two phenomena appear to be  well separated in  time and therefore it is commonly believed that they can be formally decoupled \cite{dupe2013mechanism}.  We can then, following \cite{srinivasan2002mechanism}, neglect the Landau-type component of the transformation by associating the  primary order parameter  with the shuffle.

 Note next that a classical shuffle mode  would  have naturally  emerged  in our picture of S-T transition if in our recovery of atomistic deformation gradients we had used  a double unit cell \cite{trinkle2003theoretical}.  Usually the BCC-HCP shuffle  is perceived as  proceeding via  softening of an  optical mode with the formation of an intermediate  orthorhombic configuration.  The  idea is that  such   lowering of symmetry is maintained until   the system locks-in in the higher symmetry configuration \cite{zahn2004nucleation}.  Our analysis suggests that  instead of gradual softening, the emerging  crystallographically specific  anti-parallel shifts of consecutive planes   can be   viewed  as  a layer by layer pattern  formation inside  a single  unit-cell (with some homogeneous adjustment  layer-wise). Moreover, in the lock-in state the implied  micro-heterogeneous coexistence of different variants  of the orthorhombic phase, collectively  recovering  the HCP symmetry,  can be interpreted as a special nano-twinning with   individual twins  distinguished by a lattice invariant shear.  It is the large transformation strain  in the lock-in conditions  which drives the scale of such effective   twinning to atomic dimensions. 

The proposed  analogy should be, of course,   viewed only in a metaphoric sense  as the  actual BCC-HCP transition in 3D remains fundamentally  different from the S-T transition in 2D. Thus, it is  not clear whether  the experimentally confirmed   path  for BCC-HCP transition \cite{mao1967effect}  can be indeed decomposed into  full plastic slips    or instead represented by   alternating  stacking faults resulting from   the  passage of   partial dislocations. Interestingly, the reconstructive FCC-HCP transition appear to be  an example  of the latter possibility as in this case instead of `fully' plastic deformation we   see  the micro-deformation which can be interpreted as  only `partially' plastic. Indeed, during FCC-HCP transition the   HCP phase  appears to be emerging from  an anti-parallel coordinated  gliding  of  Shockley partials     \cite{liu2004stress,yang2006factors,liu2005thermally, mahajan1977model,singh1982mechanism}.  In  the setting of S-T transformation the implied  nano-scale  stacking fault  laminates \cite{rosa2022martensitic,jin2005crystallography}   would  correspond to the   layering of the type T$_1$- S -T$_2$. Since in our model the S phase is fully destabilized at the transformation threshold, such `partially' plastic  laminates are not observed with partials appearing only transiently as it is clear from  our Fig. \ref{fig:14}(a).

\section{Conclusions}

  We begin by stressing once again that, to the best of our knowledge,  the  proposed   method of tracking the history of atomic-scale metric tensors in MD simulations is without precedent. This methodological advance offered us a unique perspective on  micro-pattern formation during reconstructive phase transitions, resolving a  long-standing problem  of   meso-scopic interpretation of mico-scale numerical experiments. Our reliance on the purely geometrical tessellation of the configurational space of metric tensors creates for the first time the real  possibility to distinguish in MD experiments between elastic and plastic deformations. 
  
  We provided compelling evidence  that the  proposed new perspective can reveal  previously hidden details of the deformation paths of reconstructive phase transitions. The emerging new interpretation of MD data goes much beyond  the conventional reasoning in terms of gamma surfaces. In particular, it brings to the forefront for the first time   the dominant role in the  formation of polycrystalline textures   at reconstructive  phase transformations of lattice invariant shears, see also \cite{bhattacharya2004crystal}.  

Given that   reconstructive phase transitions  could not  be rationalized in the framework of the conventional Landau theory of phase transitions, our novel approach offers a compromise: we effectively interpret  lattice invariant shears as representing quasi-Landau phases. Our work can be then viewed as a  response to the challenge of the  development of a   Landau-type theory  of reconstructive transitions, offering  a  paradigm changing, plasticity-centered   interpretation of the corresponding transformation paths. The discovered slip-dominated mechanism of reconstructive phase transitions is  purely geometrical   and is therefore insensitive to microscopic details. Therefore it  can be   viewed as a robust feature of a broad class of structural transformations including the  iconic BCC-HCP and FCC-HCP transitions, where lattice invariant shears would  emerge under the disguise of microscale shuffling.  Needless to say that we expect   our theoretical predictions to  stimulate considerable experimental efforts aimed at the recovery  of the predicted nano-twinning patterns  in realistic crystals far beyond the toy  model of square to hexagonal transition.

Finally, we mention that our results  have  implications for the whole field of solid state  physics as we  build a new  bridge between the microscopic stability  of crystals   and their macroscopic mechanical behavior  usually addressed through engineering phenomenological plasticity theory. An important    theoretical advance is  the  development of a broader perspective on structural phase transitions which goes beyond the classical Landau approach.



\begin{acknowledgments}  
 The authors are grateful to D. Gratias, C. Denoual and Y.P. Pellegrini for helpful suggestions in the course of this work. K.G and O.U.S. were supported by Grant  ANR-19-CE08-0010.   L.T.  acknowledges the support of the  Grants ANR-17-CE08-0047-02, ANR–21-CE08-MESOCRYSP and ERC-H2020-MSCA-RISE-2020-101008140.
\end{acknowledgments}  



\providecommand{\noopsort}[1]{}\providecommand{\singleletter}[1]{#1}%

\end{document}